\documentclass[a4paper,11pt]{article}
\pdfoutput=1 \usepackage{jheppub_2} 
\usepackage[T1]{fontenc} 
\usepackage[utf8]{inputenc}
\usepackage{array}
\usepackage{multirow}
\usepackage{booktabs}
\usepackage{mathrsfs}
\usepackage[euler]{textgreek}

\usepackage{amsmath}
\usepackage{amsthm}
\usepackage{amsfonts}
\usepackage{amssymb}
\usepackage{graphicx}
\usepackage[font={small}]{caption}
\usepackage{float}
\usepackage[export]{adjustbox}
\usepackage[hang, flushmargin,bottom]{footmisc} 
\usepackage{hyperref}
\hypersetup{
     colorlinks   = true,
     citecolor    = blue
}
\usepackage{placeins}
\usepackage{enumitem}
\usepackage{slashed}
\usepackage{bbm}
\usepackage{appendix}

\usepackage{titlesec}
\titleformat{\chapter}[display]
  {\normalfont\LARGE\bfseries}
  {\chaptertitlename\ \thechapter}{5pt}{\LARGE}
  \titlespacing*{\chapter}{0pt}{-20pt}{35pt}
\usepackage{bigstrut}
\usepackage{pst-node}
\usepackage{pstricks}
\usepackage{datetime}
\usepackage{color, colortbl}
\definecolor{GrayLight}{gray}{0.9}

\usepackage{physics}
\usepackage{mathtools}
\usepackage{setspace}
\usepackage{fancyhdr}
\usepackage[makeroom]{cancel}
\newcommand{\be}{\begin{equation}}
\newcommand{\ee}{\end{equation}}
\newcommand{\bes}{\begin{equation*}}
\newcommand{\ees}{\end{equation*}}

\renewcommand\thesection{\arabic{section}}
\usepackage{hyperref}
\hypersetup{%
  colorlinks = true,
  linkcolor  = blue
}
\usepackage{soul}

\usepackage{placeins}
\usepackage{hyperref}
\usepackage{makeidx}
\usepackage{mathrsfs}
\usepackage{dsfont}
\usepackage{fancybox}
\usepackage{amsfonts}
\usepackage{graphicx}
\usepackage{tcolorbox}
\usepackage{marvosym} 
\usepackage{tikz}
\usepackage{tabulary}
\usepackage{pgfplots}
\usepackage{subfig}
\usepackage{amsmath}
\usepackage{slashed}
\usepackage{mathtools}
\usepackage{fancyhdr} 
\usepackage{nccmath}
\usepackage{verbatim}
\usepackage{color}
\usepackage{amsmath, amsthm, amssymb, amsfonts}
\usepackage{array}
\newcolumntype{P}[1]{>{\centering\arraybackslash}p{#1}}
\usepackage{enumerate}
\usepackage{physics}
\usepackage[font={small, up}]{caption}
\usepackage{tikz}
\usetikzlibrary{"arrows", "automata", "backgrounds", "calendar", "chains", "matrix", "mindmap", "patterns", "petri", "shadows", "shapes.geometric", "shapes.misc", "spy", "trees"}
\usepackage{appendix}
\usepackage[normalem]{ulem}

\newcommand{\imineq}[2]{\vcenter{\hbox{\includegraphics[height=#2ex]{#1}}}}

\newcommand{\myComment}[1]{}

\newcommand{\beq}{\begin{equation}}
\newcommand{\eeq}{\end{equation}}

\newcommand{\SU}{\,{\rm SU}}

\newcommand{\U}{\,{\rm U}}

\notoc 

\title{ \begin{center} Theory of Dirac Dark Matter: \\ Higgs Decays and EDMs  \end{center}}

\author[]{Pavel Fileviez P\'erez$^a$}
\author[]{and Alexis D. Plascencia$^b$}
\affiliation[a]{Physics Department and Center for Education and Research in Cosmology and Astrophysics (CERCA), 
Case Western Reserve University, Cleveland, OH 44106, USA}
\affiliation[b]{INFN, Laboratori Nazionali di Frascati, C.P. 13, 100044 Frascati, Italy}
\emailAdd{pxf112@case.edu}
\emailAdd{alexis.plascencia@lnf.infn.it}

\abstract{We discuss a simple theory predicting the existence of a Dirac dark matter candidate from gauge anomaly cancellation. In this theory, the spontaneous breaking of local baryon number at the low scale can be understood. We show that the constraint from the dark matter relic abundance implies an upper bound on the theory of a few tens of TeV. We study the correlation between the dark matter constraints and the prediction for the electric dipole moment (EDM) of the electron. We point out the implications for the diphoton decay width of the Standard Model Higgs. Furthermore, we study the decays of the new Higgs present in the theory, we show that the branching ratio into two photons can be large and discuss the correlation between the dark matter constraints and the properties of the new Higgs decays. This theory could be tested at current or future experiments by combining the results from dark matter, collider and EDM experiments.}

\begin{document} 

\maketitle


\flushbottom

\newpage 

\section{Introduction}
The possible existence of dark matter (DM) in the Universe has motivated many studies in particle physics and cosmology in order to address this problem. Currently, there is a large number of experiments looking for dark matter signatures using different approaches. This experimental program could be successful; however, the nature of dark matter remains unknown. From a theoretical point of view, we should aim to understand theories 
that predict the existence of dark matter following some well-defined theoretical principles. 

For many years the particle physics community supported the idea of having a cold dark matter candidate in the Minimal Supersymmetric Standard Model (MSSM); namely, the lightest neutralino. The MSSM could describe physics at the multi-TeV scale but in order to have a dark matter candidate, we need to impose by hand the well-known R-parity discrete symmetry, which could also be used to suppress dimension five contributions to the decay of the proton. Since this symmetry is imposed by hand, it cannot be said that the MSSM generically predicts a dark matter candidate. Unfortunately, we have a similar situation in other extensions of the Standard Model such as in theories with extra dimensions where the KK-parity is also imposed to ensure the stability of the dark matter candidate.  

Recently, we investigated simple theories~\cite{Duerr:2013dza,Perez:2014qfa} for physics beyond the Standard Model where the existence of dark matter is predicted from the cancellation of gauge anomalies. The main motivation to study these theories is the possibility to understand the spontaneous breaking of baryon number in nature if this symmetry is a local gauge symmetry as the other gauge symmetries of the SM. The field content of these theories, determined by anomaly cancellation, is very simple and as an extra feature it predicts a cold dark matter candidate. In these theories, the stability of the dark matter candidate is a natural consequence of the spontaneous breaking of local baryon number. 

In this article we investigate the simple theory for Dirac dark matter proposed in Ref.~\cite{Duerr:2013dza}. This theory provides a theoretical framework to understand the spontaneous breaking of baryon number.
In order to define an anomaly-free theory the particle content is composed of six new fermionic representations, and the dark matter is the lightest new neutral fermionic field in the theory.  For a detailed study of this theory 
see the previous studies in Refs.~\cite{Duerr:2013lka,Duerr:2014wra,FileviezPerez:2015mlm,FileviezPerez:2018jmr}. We revisit the implications coming from the relic density and direct detection constraints, 
and find an upper bound on the theory around 30 TeV. Therefore, we can hope to test the theory at current or future experiments. We study the correlation between the dark matter constraints 
and the decays of the new Higgs present in the theory. We also show that the current bounds on the SM Higgs diphoton decay width provide a non-trivial bound on the particle spectrum. 

Experimental searches for CP violation are some of the most sensitive to contributions from new physics. Namely, they are able to probe energy scales much higher than the electroweak scale whenever the CP-violating phase is large. Recently, the ACME collaboration has set a strong limit on the electron electric dipole moment (EDM)~\cite{Andreev:2018ayy},
$|d_e| / e \leq 1.1 \times 10^{-29} \, {\rm cm,}$
at the $90\%$ confidence level and they expect to improve this measurement during Stage III. For reviews on CP violation and EDMs we refer the reader to Refs.~\cite{Bernreuther:1990jx,Pospelov:2005pr,Fukuyama:2012np,Chupp:2017rkp}. 
In this article, we study the implications of the electron EDM bound in this theory. Since there is a strong upper bound on the masses of all new fermions, the EDM bounds are very important. 
We also study the correlation between the dark matter constraints, Higgs bounds and EDM experimental limits. For a study of EDMs in the context of local baryon number and Majorana dark matter see Ref.~\cite{Perez:2020jyg}.

This article is organized as follows: In Section~\ref{sec:theory} we briefly review the theory of local baryon number predicting a Dirac dark matter candidate. In Section~\ref{sec:DM} we study the phenomenology of Dirac dark matter, focusing on the upper bound that comes from not overproducing the dark matter relic density and the perturbativity of the couplings. In Section~\ref{sec:HD} we study how the new fermions modify the diphoton decay width of the SM Higgs boson. We also study the tree-level and loop-induced decays of the new Baryonic Higgs. In Section~\ref{sec:EDM} we present the study of CP violation and the implications for the EDM of the electron. Our main results are summarized in Section~\ref{sec:summary}. 

For completeness, we present different appendices with the details of the calculations performed in this work. Appendix~\ref{sec:appFR} contains the complete Feynman rules of the theory, Appendix~\ref{sec:diagonal} has the diagonalization of the mass matrices, in Appendix~\ref{sec:appCP} we present a discussion on the CP-violating phases and in Appendix~\ref{sec:EDMcalc} we present the different contributions to the EDMs. In Appendix~\ref{sec:appDecaysHB} we provide the full expressions for the tree-level and loop-induced decay widths of the Baryonic Higgs and  Appendix~\ref{sec:appDecays} has the loop functions involved in these decays and also in the Higgs diphoton decay.

\section{Simple Theory for Dirac Dark Matter}
\label{sec:theory}
%
\begin{table}[t]\setlength{\bigstrutjot}{6pt}
\centering
\begin{tabular}{|ccccc|}\hline
Fields & $\SU(3)_C$ & $\SU(2)_L$ & $\U(1)_Y$  & $\U(1)_B$ \bigstrut\\ \hline \hline
$\Psi_L = \mqty(\Psi_L^0 \\[0.5ex] \Psi_L^-)$ & $\mathbf{1}$ & $\mathbf{2}$ & $ -\frac{1}{2}$ & $B_1$\bigstrut\\[3ex]
$\Psi_R = \mqty(\Psi_R^0 \\[0.5ex] \Psi_R^-)$ & $\mathbf{1}$ & $\mathbf{2}$ & $-\frac{1}{2}$ & $B_2$  \bigstrut\\
$\eta_R$ & $\mathbf{1}$ & $\mathbf{1}$ & $-1$ & $B_1$ \bigstrut\\
$\eta_L$ & $\mathbf{1}$ & $\mathbf{1}$ & $-1$ & $B_2$ \bigstrut\\
$\chi_R$ & $\mathbf{1}$ & $\mathbf{1}$ & $0$ & $B_1$ \bigstrut\\
$\chi_L$ & $\mathbf{1}$ & $\mathbf{1}$ & $0$ & $B_2$ \bigstrut\\
\hline
  \end{tabular}
  \caption{Fermionic representations needed for gauge anomaly cancellation with $B_1 -  B_2 = -3$~\cite{Duerr:2013dza}.}
  \label{tabla1}
\end{table}
The origin of baryon number violation in nature is unknown. The theory proposed in Ref.~\cite{Duerr:2013dza} provides a way to understand the spontaneous violation of baryon number in nature.
This theory is based on the gauge symmetry $$\SU(3)_C \otimes \SU(2)_L \otimes \U(1)_Y \otimes \U(1)_B,$$ 
where the extra Abelian symmetry, $\U(1)_B$, corresponds to local baryon number. In order to study the spontaneous breaking of local baryon number we need to define an anomaly-free theory.
In Table~\ref{tabla1} we list the extra fermionic content needed to cancel all the gauge anomalies. Notice that the extra fermions have baryon numbers $B_1$ or $B_2$, but anomaly cancellation imposes 
the condition $B_1-B_2=-3$. The full Lagrangian of this theory can be written as
\begin{eqnarray}
 \mathcal{L} & = &  \mathcal{L}_{\rm SM} - \frac{g_B}{3} ( \bar{Q}_L \gamma^\mu Q_L +  \bar{u}_R \gamma^\mu u_R + \bar{d}_R \gamma^\mu d_R ) Z_\mu^B \nonumber \\
& -& \frac{1}{4} Z_{\mu \nu}^B Z^{B, \mu \nu} 
 +  \mathcal{L}_{K}^B +  \mathcal{L}_{Y}^B - V(H,S_B), 
\end{eqnarray}
where $ \mathcal{L}_{\rm SM}$ is the SM Lagrangian, $Z_\mu^B$ is the leptophobic gauge boson associated to$\,\U(1)_B$, and  $Z_{\mu \nu}^B = \partial_\mu Z_\nu^B - \partial_\nu Z_\mu^B$.
In the above equation $Q_L \sim (\mathbf{3},\mathbf{2},1/6,1/3)$, $u_R \sim (\mathbf{3},\mathbf{1},2/3,1/3)$ and $d_R \sim (\mathbf{3},\mathbf{1},-1/3,1/3)$ are the multiplets for the Standard Model quarks.
Here we are neglecting the kinetic mixing between the Abelian symmetries and assuming $B_1 \neq - B_2$ in order to avoid the case with Majorana dark matter that has been already studied in Ref.~\cite{FileviezPerez:2019jju}.

The new kinetic terms are given by
\begin{eqnarray}
 \mathcal{L}_{K}^B & = & i \bar \Psi_L  \slashed{D} \Psi_L + i \bar \Psi_R  \slashed{D} \Psi_R +  i \bar \eta_R  \slashed{D} \eta_R  + i \bar \eta_L  \slashed{D} \eta_L \nonumber \\
& + &  i \bar \chi_R  \slashed{D} \chi_R +  i \bar \chi_L  \slashed{D} \chi_L + (D_\mu S_B)^\dagger (D^\mu S_B), 
\end{eqnarray} 
while the new Yukawa interactions can be written as
\begin{eqnarray}
\label{eq:Yukawas}
- \mathcal{L}_{Y}^B & = & y_1 \bar \Psi_L H \eta_R + y_2 \bar \Psi_R H  \eta_L + y_3 \bar \Psi_L \tilde{H} \chi_R + y_4 \bar \Psi_R \tilde{H} \chi_L \nonumber \\
& + &  y_\Psi \bar \Psi_L \Psi_R S_B^* + y_\eta \bar \eta_R \eta_L S_B^* + y_\chi \bar \chi_R \chi_L S_B^*  +  \rm{h.c.} \, ,
\end{eqnarray}
where $H\sim(\mathbf{1}, \mathbf{2}, 1/2,0)$ is the Standard Model Higgs, and $\tilde{H}=i\sigma_2 H^*$. The scalar $S_B \sim (\mathbf{1},\mathbf{1},0,3)$ is responsible for the spontaneous breaking of baryon number.
We define the following mass parameters,
\beq
\mu_\Psi = \frac{y_\Psi}{\sqrt{2}} v_B, \hspace{1cm} \mu_\eta=\frac{y_\eta}{\sqrt{2}} v_B, \hspace{1cm} \mu_\chi=\frac{y_\chi}{\sqrt{2}} v_B, 
\eeq
and in Appendix~\ref{sec:diagonal} we discuss the diagonalization of the mass matrices to obtain the physical states. 
These interactions can generate large masses (above the electroweak scale) for the new fermions. This happens after $S_B$ acquires a non-zero vacuum expectation value (vev) and spontaneously breaks the $\U(1)_B$ symmetry. From Eq.~\eqref{eq:Yukawas} we see that $S_B$ must carry baryon number equal to three. Therefore, local baryon number must be broken in three units and the theory predicts the proton to be stable. Since the proton is stable in this theory, it can describe physics at the low 
scale in agreement with experimental constraints.

The full scalar potential is given by
\begin{eqnarray}
V &=& - \mu_H^2 H^\dagger H + \lambda_H (H^\dagger H)^2 
- \mu_B^2 S_B^\dagger S_B + \lambda_B (S_B^\dagger S_B)^2  + \lambda_{HB} (H^\dagger H) (S_B^\dagger S_B).
\end{eqnarray}
The Higgs fields can be written as
\begin{equation}
H = \mqty(G^+ \\[0.5ex] \frac{1}{\sqrt{2}} (v_0 + h_0 + i G_0)), \hspace{0.5cm} { \rm{and}} \hspace{0.5cm} S_B = \frac{1}{\sqrt{2}} (v_B + s_B + i G_B),
\end{equation}
with the physical Higgs bosons defined as follows
\begin{equation}
\begin{split}
	h &= h_0 \cos{\theta_B} - s_B \sin{\theta_B},\\
	h_B &= s_B \cos{\theta_B} + h_0 \sin{\theta_B},
	\end{split}
\end{equation}
where $\theta_B$ corresponds to the mixing angle that diagonalizes the mass matrix for the Higgs bosons and it is given by
\begin{align}
	\tan{2 \theta_B} = \frac{v_0 v_B \lambda_{HB}}{ v_B^2 \lambda_B - v_0^2 \lambda_H },
\end{align}
where $v_0$ and $v_B$ correspond to the vevs of $H$ and $S_B$, respectively.
It is important to mention that this simple theory predicts only few extra physical fields: 
\begin{itemize}

\item Two neutral Dirac fermions, $\chi_i^0$,  with $i=1,2$.

\item Two electrically charged fermions $\chi_j^{\pm}$ with $j=1,2$.

\item  A new Higgs, $h_B$, called Baryonic Higgs.

\item The spin one gauge boson, $Z_B$, associated to$\,\U(1)_B$. 

\end{itemize}
After symmetry breaking there is an anomaly-free global symmetry in the new sector:

\begin{equation}
\chi_L \to e^{i \theta} \chi_L, \ \chi_R \to e^{i \theta} \chi_R, \  \Psi_L \to e^{i \theta} \Psi_L, \  \Psi_R \to e^{i \theta} \Psi_R, \  \eta_L \to e^{i \theta} \eta_L, \  \eta_R \to e^{i \theta} \eta_R,
\end{equation}
that is remnant from the local baryon symmetry. Therefore, the lightest new field in this sector is stable. The stable field in this sector should be neutral to avoid issues with cosmology 
and then we have a candidate to describe the cold dark matter in the Universe. It is possible to consider two different scenarios for the lightest field: a) $\chi_1^0$ is $\Psi$-like, b) $\chi_1^0$ is $\chi$-like.
The first scenario is ruled out by direct detection experiments because $\Psi^0$ is a $\SU(2)_L$ doublet and the cross-section is several orders of magnitude above the experimental limit~\cite{FileviezPerez:2018jmr}.
We focus on the second scenario, which is consistent with all the experimental limits, as we will discuss below.

\section{Dirac Dark Matter}
\label{sec:DM}
\begin{figure}[t]
\centering
\includegraphics[width=0.65\linewidth]{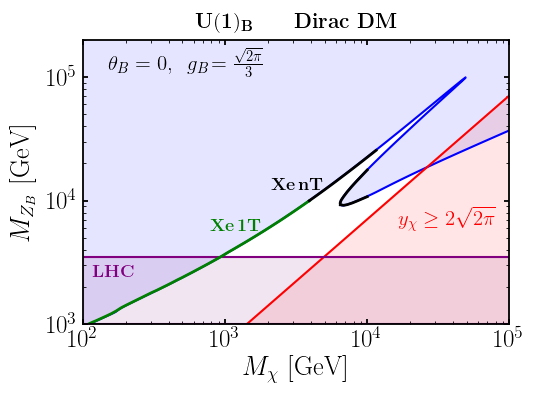}
\includegraphics[width=0.65\linewidth]{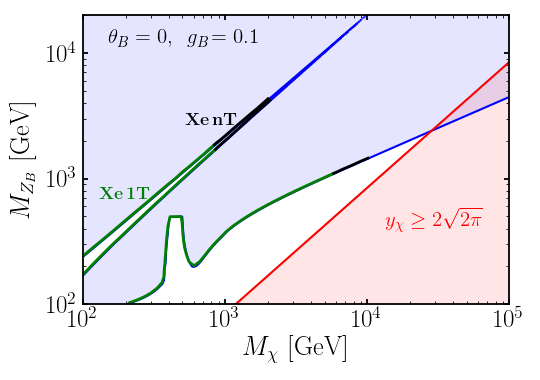}
\caption{Dark matter relic density in the $M_{Z_B}$ vs $M_\chi$ plane. The solid blue line reproduces the measured value of $\Omega_{\rm DM} h^2\!=\!0.12$~\cite{Planck:2018vyg} while the region shaded in light blue overproduces the DM relic density. The region shaded in purple is excluded by dijet resonance searches at the LHC~\cite{ATLAS:2017eqx,CMS:2018mgb}. The region shaded in red is excluded by the perturbativity of the Yukawa coupling $y_\chi$. The solid green line is excluded by Xenon-1T~\cite{XENON:2018voc} while the solid black line shows the projected sensitivity for Xenon-nT~\cite{XENON:2015gkh}. The upper panel corresponds to the maximal coupling $g_B=\sqrt{2\pi}/3$, while the lower panel is for $g_B=0.1$. In both panels we fixed $M_{h_B}=1$ TeV, $\theta_B=0$ and the baryonic charge to $B_1\!=\!-1/2$ which implies $B_2\!=\!5/2$.}
\label{fig:DM}
\end{figure}
This theory predicts generically a Dirac dark matter candidate that is $\chi$-like. Here $\chi=\chi_L + \chi_R$ is the DM candidate and the relevant interactions are given by
\begin{eqnarray}
\mathcal{L} && \supset \frac{1}{3} g_B \bar{q} \gamma^\mu q Z_\mu^B - g_B \bar{\chi} \gamma^\mu \left(  B_2 P_L + B_1 P_R \right) \chi Z_\mu^B \nonumber \\
&&- M_\chi \bar{\chi} \chi + \frac{M_\chi \cos \theta_B}{v_B} \bar{\chi} \chi h_B - \frac{M_\chi \sin \theta_B}{v_B} \bar{\chi} \chi h,
\end{eqnarray}
where $M_\chi=y_\chi v_B/\sqrt{2}$ and $P_{L(R)}=(1\mp \gamma_5)/2$. 

In order to study the relic density constraints we need to consider all possible annihilation channels:
$$\bar{\chi} \chi \to \bar{q} q, \,\,\, Z_B Z_B, \,\,\, h Z_B, \,\,\, h_B Z_B, \,\,\, h h, \,\,\, h h_B, \,\,\, h_B h_B, \,\,\, WW, \,\,\, ZZ.$$
The constraints from DM direct detection are very important to understand the allowed parameter space in this theory. The elastic nucleon-$\chi$ cross-section has two main contributions mediated by the $Z_B$ gauge boson and the Higgs bosons.

In Fig.~\ref{fig:DM} we present our results for the calculation of the dark matter relic density which has been computed numerically using \texttt{MicrOMEGAs 5.0.6} \cite{Belanger:2018ccd}. The solid blue line reproduces the measured value of $\Omega_{\rm DM} h^2\!=\!0.12$~\cite{Planck:2018vyg} while the region shaded in blue overproduces the dark matter relic abundance. In the upper panel we fixed the gauge coupling to its largest value allowed by perturbativity $g_B=\sqrt{2\pi}/3$. The red line corresponds to the maximal value for the Yukawa coupling allowed by perturbativity; namely, $y_\chi=2\sqrt{2\pi}$. We find that the results for the relic density are almost independent of the scalar mixing angle; for the plots we set $\theta_B=0$. The peak that can be observed in the plots corresponds to the resonance $M_\chi \simeq M_{Z_B}/2$ so that $\chi \bar{\chi} \to q \bar{q}$ is the dominant annihilation channel. The bump in the lower part of the plot also corresponds to a resonance $M_\chi \simeq M_{h_B}/2$. In the entire region above the TeV scale the dominant annihilation channel is $\chi \bar{\chi} \to Z_B h_B$ and this does not require being close to any resonance; therefore it is the generic DM annihilation channel.

In contrast to the case with Majorana dark matter studied in Ref.~\cite{FileviezPerez:2019jju}, in this scenario the direct detection cross-section does not have velocity suppression, and hence, these bounds are much stronger. In Fig.~\ref{fig:DM} we show with a solid green line the region that is excluded by Xenon-1T~\cite{XENON:2018voc} while the solid black line shows the projected sensitivity for Xenon-nT~\cite{XENON:2015gkh}. This is discussed in more detail in Ref.~\cite{FileviezPerez:2018jmr} where a study of the Dirac dark matter phenomenology was performed. As it was shown in that work, the experimental bound from Xenon-1T rules out multi-TeV dark matter masses even in the case of zero mixing with the SM Higgs. Furthermore, the projected sensitivity for Xenon-nT will be able to probe a large region in the parameter space.

The region shaded in purple in Fig.~\ref{fig:DM} shows the parameter space excluded by dijet resonance searches at the LHC~\cite{ATLAS:2017eqx,CMS:2018mgb}. In the lower panel of that figure we show the results for $g_B=0.1$, in this scenario the gauge boson can evade the constraints from the dijet searches. However, the experimental constraint from direct detection requires $M_\chi \gtrsim 850$ GeV if the dark matter density is saturated.

The upper panel in Fig.~\ref{fig:DM} also shows that there is an upper bound on the masses of the gauge boson and the dark matter in order to not overclose the Universe. In summary, we obtain $M_{Z_B}\lesssim 100$ TeV and $M_\chi \lesssim 50$ TeV as upper bounds for the masses of the gauge boson and the dark matter, respectively. This bound is stronger than the one coming from the unitarity of the S-matrix which is around 200 TeV~\cite{FileviezPerez:2018jmr}. Furthermore, ignoring the resonant region, which requires $M_\chi\approx M_{Z_B}/2$, the upper bounds become $M_{Z_B}\lesssim 19$ TeV and $M_\chi \lesssim 26$ TeV which are more generic.

Furthermore, since all the new fermions acquire their mass from the $\U(1)_B$ breaking scale, there is a non-decoupling effect within the new sector and the upper bound also applies to the charged fields that contribute to the EDMs.
Namely, in the limit of small fermionic mixing we have that $M_{\chi_1^\pm}\!\simeq\!y_\eta v_B/\sqrt{2} \!=\!y_\eta M_{Z_B}/(3\sqrt{2} g_B)$ and $M_{\chi_2^\pm}\!\simeq\!y_\Psi v_B/\sqrt{2}$, and hence, setting the Yukawa couplings to their allowed values by perturbativity $y_\Psi, \, y_\eta \leq 2\sqrt{2\pi}$ we find that,
\beq
M_{\chi_i^\pm} \lesssim 140 \, \, {\rm TeV} \,\,\, (30 \, \, {\rm TeV}),
\eeq
where the number in parentheses is more generic since it does not require the theory to live in a resonance.
This is a striking result that implies that we can hope to fully test this theory at current or future collider experiments\footnote{This upper bound is obtained from the largest value of $g_B$ allowed by perturbativity. However, due to the positive $\beta$-function for $g_B$, such a large value will quickly run into a Landau pole, and hence, we expect the theory to live at a much lower scale.} .

\section{Higgs Decays}
\label{sec:HD}

In 2012 both the ATLAS and CMS announced the discovery of a scalar particle with properties similar to the SM Higgs boson with a mass of 125 GeV. The presence of new physics coupled to this particle motivates a detailed experimental study of its properties. The diphoton decay of the Higgs represents one of its cleanest signatures and its measurement is expected to be improved by the high luminosity run of the LHC. In this section, we study how the new charged fermions could modify this branching ratio. Furthermore, the theory also predicts a second Higgs that mixes with the SM Higgs, with mixing angle $\theta_B$, and we study its decays channels in detail including the loop-induced decays with the anomaly-canceling fermions running in the loop.

\subsection{SM Higgs Diphoton Decay}
\label{sec:haa}
\begin{figure}[b]
\centering
\includegraphics[width=0.75\linewidth]{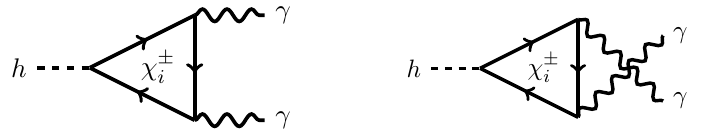}
\caption{Feynman diagrams for the contribution of the new fermions to the $h \to \gamma \gamma$ decay.}
\label{fig:feynmandiagrams}
\end{figure}

The presence of new physics coupled to the SM Higgs boson can modify some of its properties. In this theory, the new charged fermions introduced to cancel the gauge anomalies give new contributions to the Higgs decay into two photons. In Fig.~\ref{fig:feynmandiagrams} we present the Feynman diagrams that contribute to the $h\to \gamma \gamma$ decay. The signal strength for this channel normalized with respect to the SM prediction has been measured by the ATLAS collaboration to be $\mu_{\gamma \gamma}=1.16\pm 0.14$~\cite{ATLAS:2018hxb}, while the CMS collaboration reports $\mu_{\gamma \gamma}=1.12\pm 0.09$~\cite{CMS:2021kom}. The high luminosity run at the LHC is expected to improve the measurement of this rate by a factor of two~\cite{CMS:2013xfa} and a possible deviation from the SM prediction could be observed. 

There are different extensions of the SM that lead to a modification of the Higgs diphoton rate, for studies from an EFT perspective see e.g. Refs.~\cite{McKeen:2012av,Korchin:2013ifa,Alloul:2013naa,Chen:2014gka,ATLAS:2015yrd}.
For phenomenological studies of models with new vector-like fermions that have Yukawa interactions with the SM Higgs see e.g. Refs.~\cite{Voloshin:2012tv,Altmannshofer:2013zba,Chao:2014dpa,Bizot:2015zaa}.
\begin{figure}[t]
\centering
\includegraphics[width=0.495\linewidth]{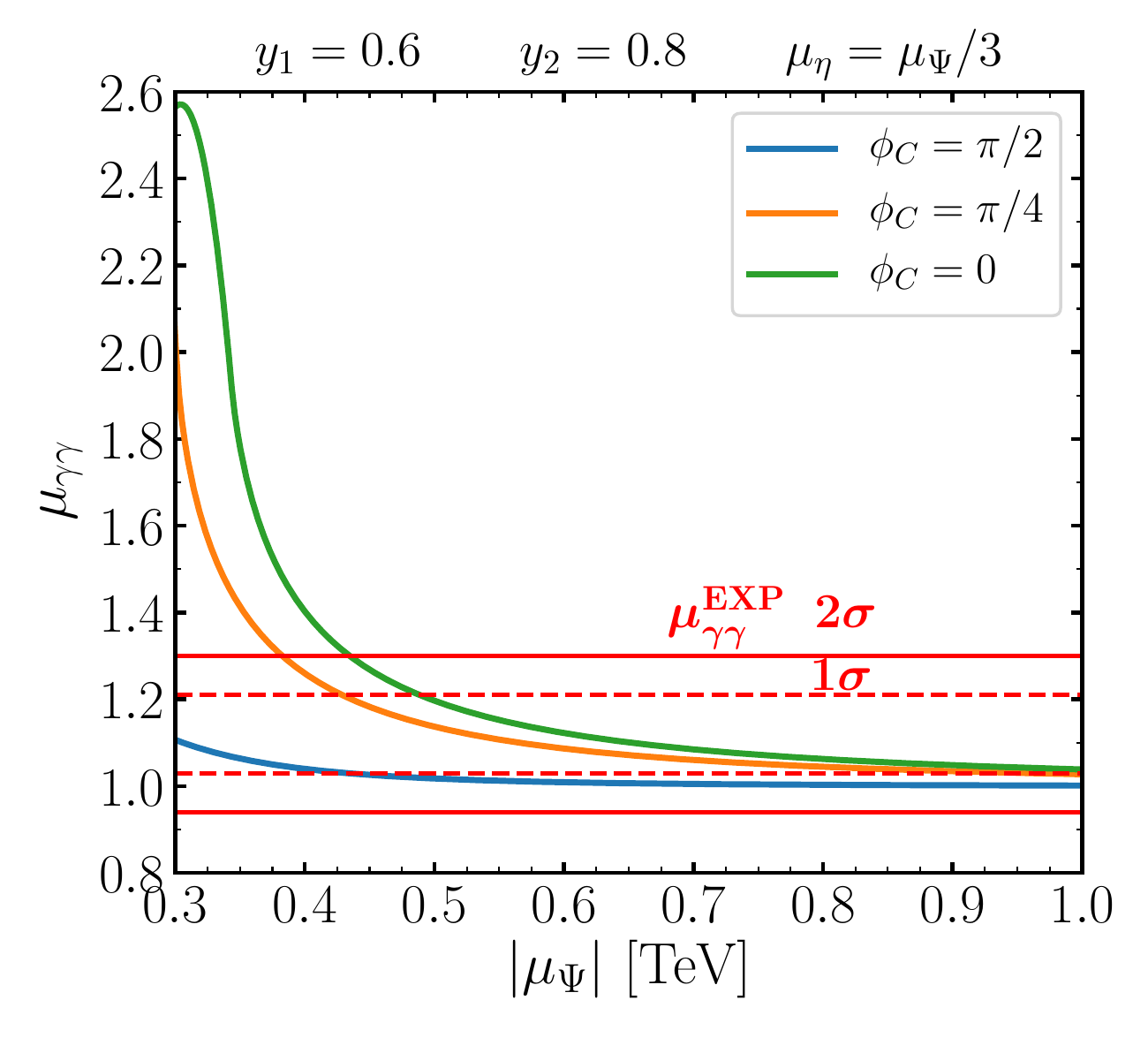}
\includegraphics[width=0.495\linewidth]{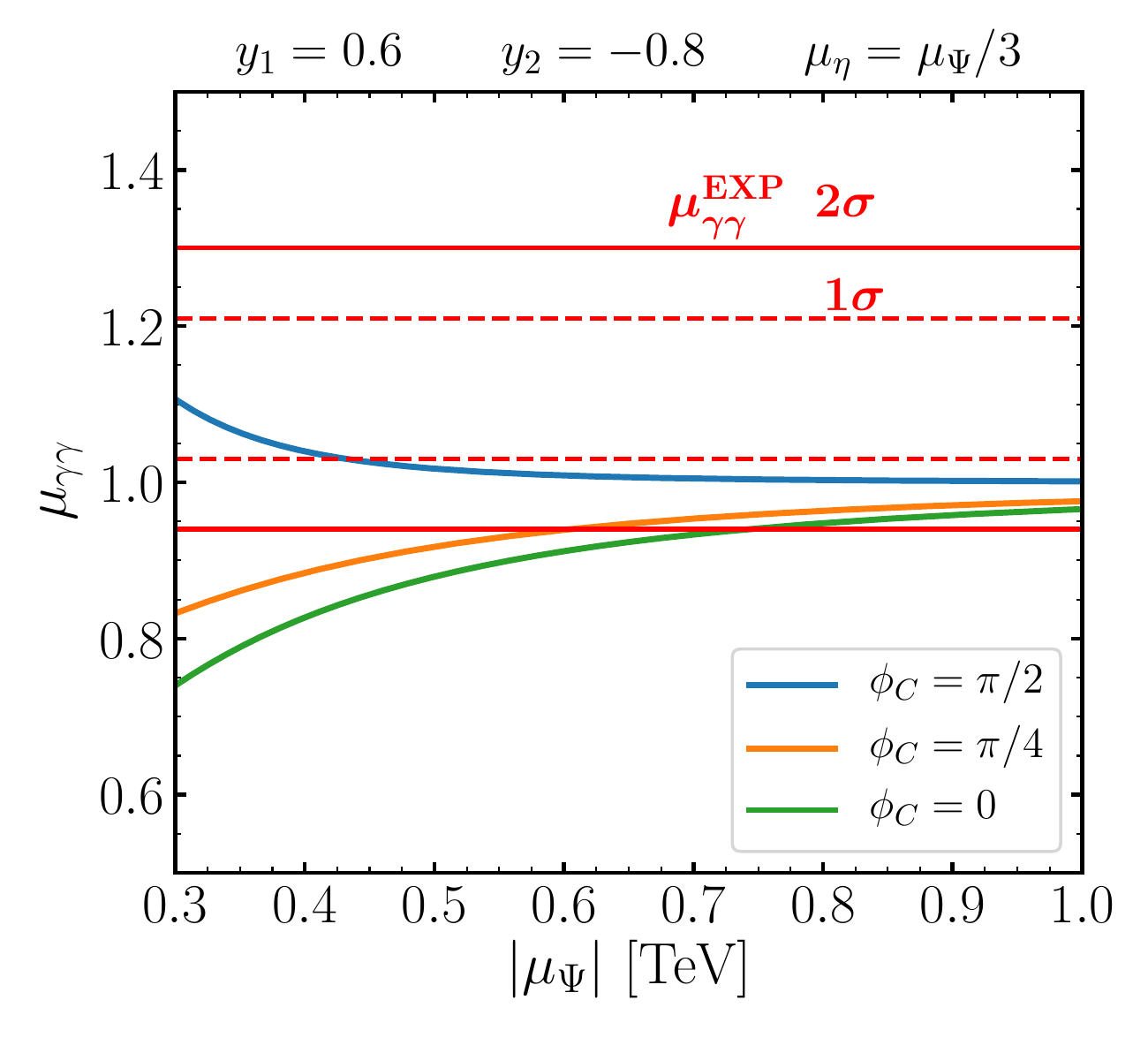}
\caption{Diphoton signal strength of the SM Higgs boson $\mu_{\gamma \gamma}$ as a function of the mass parameter $|\mu_\Psi|$ in units of TeV. The region within the solid (dashed) red lines shows the measurement by the CMS collaboration within $2\sigma$ ($1\sigma$)~\cite{CMS:2021kom}. The green, orange and blue solid lines correspond to different values for the CP-violating phase $\phi_C=0$, $\phi_C=\pi/4$ and $\phi_C=\pi/2$, respectively. }
\label{fig:diphoton}
\end{figure}
In this theory, the Higgs diphoton decay width is given by,
\begin{align}
\Gamma(h \to \gamma \gamma) = &  \frac{\alpha^2}{64  \pi^3  M_{h}^5} \left | \sum_{i=1}^2  M_{\chi^\pm_i} {\rm Re}[C_{hc}^{ii}] \, F_{\chi^\pm_i} + \frac{\cos \theta_B}{v_0} \left( \sum_{f^+} N_c^f Q_f^2 m_{f^+}^2 F_{f^+} -   F_W \right) \right |^2  \nonumber \\
& + \frac{\alpha^2}{64  \pi^3  M_{h}} \left|  \sum_{i=1}^2  M_{\chi_i^\pm} \, {\rm Im}[C_{hc}^{ii}] \, G_{\chi_i^\pm} \right|^2,  \label{eq:haa}
\end{align}
where the sum over $f^+$ is over the SM fermions, $v_0$ corresponds to the vev of the SM Higgs and the loop functions are given in Appendix~\ref{sec:appDecays}. The CP-even part of the new Yukawa interactions (real part) interfere with the SM contribution, and hence, the rate for $h \to \gamma \gamma$ can be enhanced or suppressed depending on the sign of the new couplings~\cite{Voloshin:2012tv}. The CP-odd part of the new interactions (imaginary part) does not interfere with the SM contribution and always enhances this rate; however, it only gives a small contribution. Consequently, this observable will give stronger constraints for small CP-violating phases and will be complementary to the bounds from EDMs.

As we discuss in Appendix~\ref{sec:appCP} the theory contains two independent CP-violating phases: $\phi_C={\rm arg}(y_\eta y_\Psi^* y_1 y_2^*)$ in the charged sector and $\phi_N={\rm arg}(y_\chi y_\Psi^* y_3 y_4^*)$ in the neutral sector. In order to make predictions we need to fix some of the parameters, throughout this article we fix the Yukawa couplings $y_i$ to different $\mathcal{O}(1)$ values in order to show different slices of the parameter space. We also require the charged fermion masses to satisfy the lower bound from LEP of 90 GeV~\cite{Egana-Ugrinovic:2018roi}.

In Fig.~\ref{fig:diphoton} we show our results for the Higgs diphoton signal strength $\mu_{\gamma\gamma}\equiv \sigma^{\rm prod} {\rm Br}(h \to \gamma \gamma)/( \sigma^{\rm prod}_{\rm SM} {\rm Br}(h \to \gamma \gamma)_{\rm SM})$ as a function of the mass parameter $|\mu_\Psi|$. The region within the solid (dashed) red lines shows the measurement by the CMS collaboration within $2\sigma$ ($1\sigma$). On the left panel we fix the parameters to $y_1=0.6$, $y_2=0.8$ and $\mu_\eta=\mu_\Psi/3$. The green line corresponds to $\phi_C=0$ and using the $2\sigma$ range for $\mu_{\gamma\gamma}^{\rm EXP}$ we find that $|\mu_\Psi|<0.43$ TeV is excluded. The orange line corresponds to $\phi_C=\pi/4$ for which $|\mu_\Psi|<0.38$ TeV is excluded. Finally, the blue line is for $\phi_C=\pi/2$ for which we find that the measurement of $\mu_{\gamma\gamma}$ provides no constraint. 

On the right panel in Fig.~\ref{fig:diphoton} we fix $y_1=0.6$ and $y_2=-0.8$ from which it is possible to obtain a destructive interference and lower the prediction for $\mu_{\gamma\gamma}$. The green line corresponds to $\phi_C=0$ and using the $2\sigma$ range for $\mu_{\gamma\gamma}^{\rm EXP}$ we find that $|\mu_\Psi|<0.74$ TeV is excluded. The orange line corresponds to $\phi_C=\pi/4$ for which $|\mu_\Psi|<0.6$ TeV is excluded. Finally, the blue line is for $\phi_C=\pi/2$ for which we find again that the measurement of $\mu_{\gamma\gamma}$ provides no constraint. 

We note that even though the addition of large Yukawa couplings with the Higgs could push the electroweak vacuum to the instability region at a high scale, this can be easily addressed by the positive contribution from the Higgs portal coupling $\lambda_{HB}$ between the Higgs and the scalar responsible for the breaking of baryon number. Regarding the CP-violating phase in the neutral sector, $\phi_N$, this parameter does not enter at one-loop in the calculation of $h \to \gamma \gamma$, and hence, it is not constrained by this observable.

\begin{figure}[b]
\centering
\includegraphics[width=0.8\linewidth]{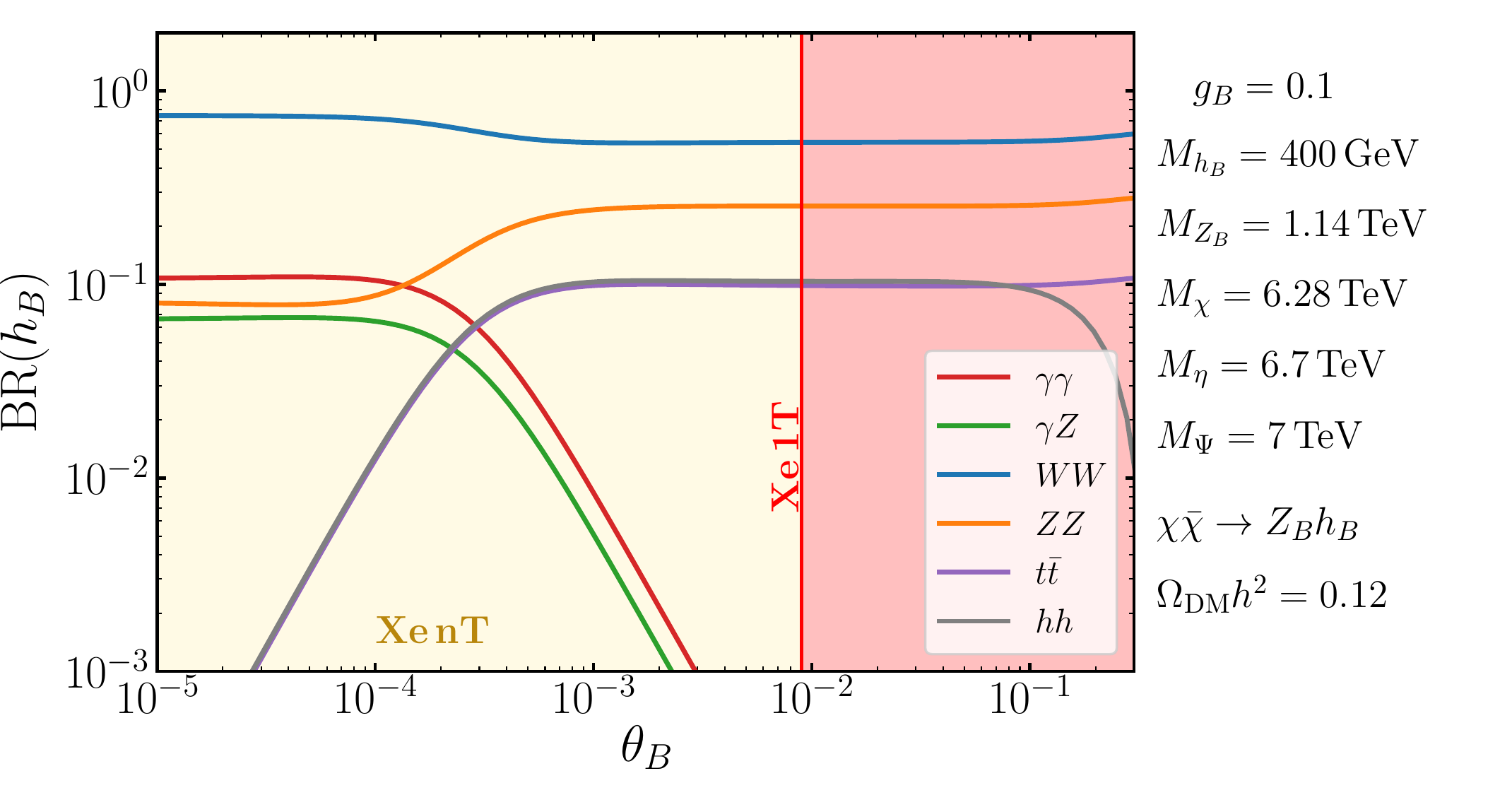} 
\caption{Branching ratios of the Baryonic Higgs $h_B$ as a function of the scalar mixing angle $\theta_B$. The relevant parameters have been fixed  to reproduce the dark matter relic density $\Omega_{\rm DM} h^2 = 0.12$ as indicated in the plot, the dominant DM annihilation channel is $\chi \bar{\chi} \to Z_B h_B$. The different colors correspond to different decay channels as shown in the plot. The area shaded in red is ruled out by Xenon-1T direct detection constraints, while in yellow we show the region that will be probed by Xenon-nT.}
\label{fig:BR1}
\end{figure}

The Yukawa couplings $y_1$ and $y_2$ will induce a mass splitting among the components of the $\SU(2)_L$ doublet $\Psi$, and hence, there will be a contribution to the electroweak precision observables $S$, $T$ and $U$. This was studied for a similar scenario in Ref.~\cite{FileviezPerez:2011pt}. Nevertheless, there exists freedom in the couplings $y_3$ and $y_4$ that also contribute to the mass splitting and those bounds can be easily satisfied. Furthermore, if the dark matter relic density is saturated, the bound from direct detection requires the dark matter mass to be above 850 GeV, and hence, the new charged fermions will be above this value so the contribution to those parameters is further suppressed.

\subsection{Baryonic Higgs Decays}
\label{sec:hb}

In this section, we study the phenomenology of the scalar responsible for the spontaneous breaking of $\U(1)_B$. As we have discussed in Section~\ref{sec:DM} the strong constraints from DM direct detection experiments require the new gauge boson and the dark matter to be above the TeV scale. Therefore, if the Baryonic Higgs has a mass around the electroweak scale it will decay dominantly into SM gauge bosons. In this section we neglect the effect of CP violation. 

\begin{figure}[b]
\centering
\includegraphics[width=0.8\linewidth]{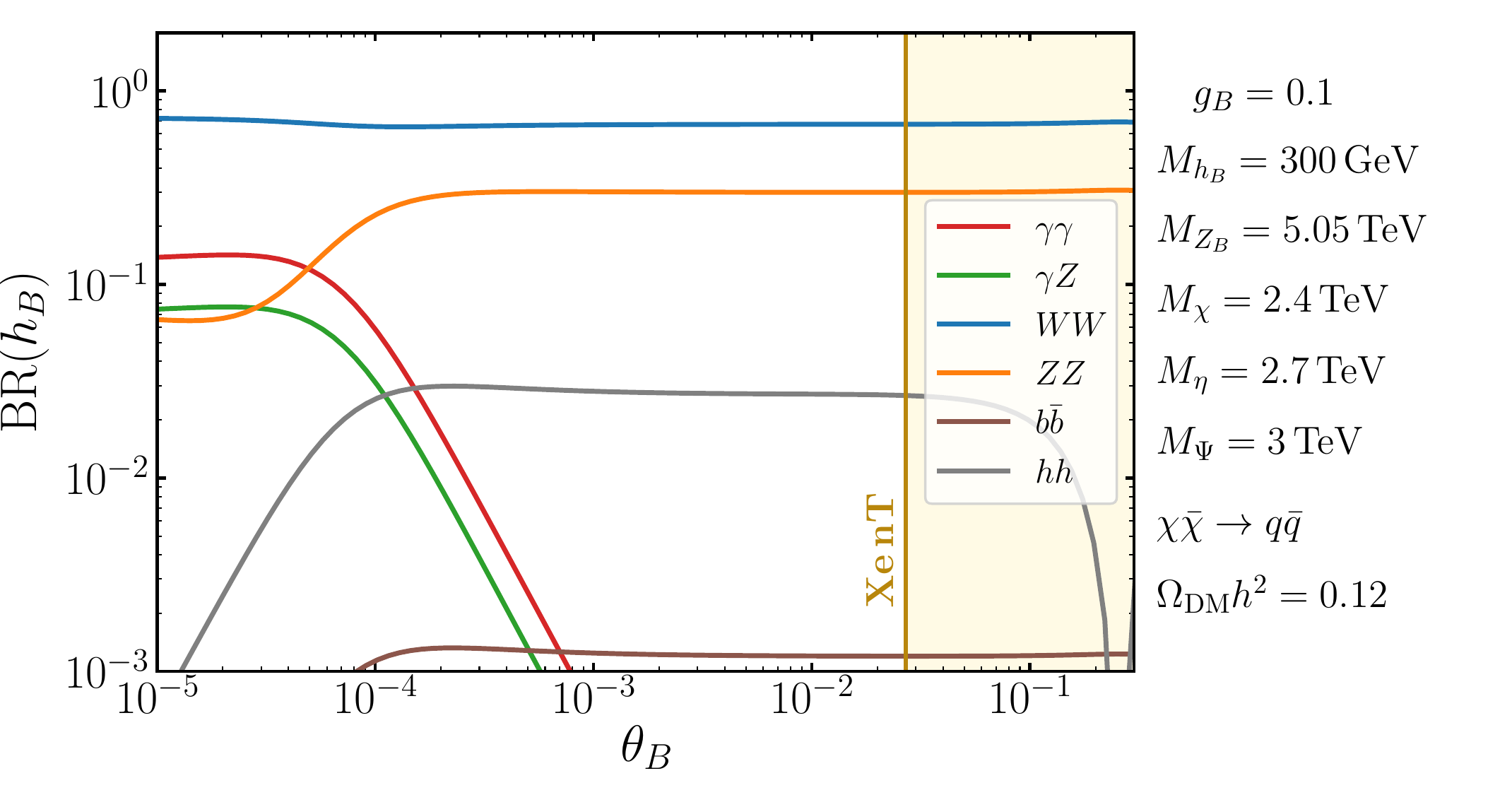} 
\caption{Branching ratios of the Baryonic Higgs $h_B$ as a function of the scalar mixing angle $\theta_B$. The relevant parameters have been fixed  to reproduce the dark matter relic density $\Omega_{\rm DM} h^2 = 0.12$ as indicated in the plot, the dominant DM annihilation channel is $\chi \bar{\chi} \to q \bar{q}$. The different colors correspond to different decay channels as shown in the plot. The area shaded in yellow shows the projected sensitivity for Xenon-nT.}
\label{fig:BR2}
\end{figure}

From the Yukawa interactions in Eq.~\eqref{eq:Yukawas} it can be seen that the anomaly-canceling fermions running in the loop will induce decays of $h_B$ into SM gauge bosons including $\gamma \gamma$. We computed these decay widths with the help of Package-X~\cite{Patel:2015tea} and present the full expressions in Appendix~\ref{sec:appDecaysHB}. Of particular interest is the clean diphoton channel that, as we will show, can have a branching ratio much larger than the one for the SM Higgs.

In Fig.~\ref{fig:BR1} we show our results for the branching ratio of $h_B$ as a function of the scalar mixing angle $\theta_B$. As we have discussed in Sec.~\ref{sec:DM} the most generic case is for $\chi \bar{\chi} \to Z_B h_B$ to be the dominant DM annihilation channel. Consequently, we fix the parameters to be in this regime and reproduce the measured relic abundance of $\Omega_{\rm DM} h^2 = 0.12$. The experimental bound from Xenon-1T is shaded in red and constrains the mixing angle to be $\theta_B \lesssim 0.009$, while the projected sensitivity for Xenon-nT will be able to probe the $Z_B$ mediated cross-section, and hence, will be sensitive to a vanishing scalar mixing angle. For the whole range of mixing angles the branching ratio for the decay into a pair of $Z$ bosons is large, and hence, the decay channel $h_B \to ZZ\to 4\ell$ can be used to search for this new scalar. Moreover, for small values of the mixing angle $\theta_B \leq 10^{-4}$ the branching ratio for the loop-induced decay $h_B \to \gamma \gamma$ can be as large as $10\%$. Finally, since we fix $M_{h_B} = 400$ GeV for large values of the mixing angle there is a large branching ratio into a top anti-top quark pair.

In Fig.~\ref{fig:BR2} we present our results for a scenario in which the dominant contribution to the relic density comes from the annihilation channel $\chi \bar{\chi} \to q \bar{q}$ which needs to be very close to the resonance $M_\chi\approx M_{Z_B}/2$ to reproduce the measured value of the relic abundance. In this scenario there is no constraint from Xenon-1T; nonetheless, the projected sensitivity for Xenon-nT will be able to put the constrain of $\theta_B \lesssim 0.027$ on the mixing angle. As can be seen, the dominant decays are $h_B \to W^+W^-$ and $h_B\to ZZ$. Furthermore, for very small values of the mixing angle $\theta_B \leq 10^{-4}$ then the branching ratio $h_B\to \gamma \gamma$ can be as large as $13\%$.

There are two main channels for the production of $h_B$ at the LHC. On the one hand we have the associated production $pp \to Z_B^* \to Z_B h_B$ which was studied recently in Ref.~\cite{Perez:2020baq}. Even if the constraints from DM direct detection require that the mass of the $Z_B$ has to be above the TeV, the production cross-section for $M_{h_B}=400$ GeV and $M_{Z_B}=1.1$ TeV corresponds to $\sigma(pp \to Z_B h_B)\approx 10^{-2}$ fb for a center-of-mass energy of 14 TeV, and hence, $\mathcal{O}(10)$ events can be generated in the high luminosity run at the LHC. On the other hand, there is production via gluon fusion which is suppressed by the mixing angle $\sin^2\theta_B$; however, the branching ratio for $h_B \to \gamma \gamma$ can be $10^2$ times larger than the one for the SM Higgs. Consequently, for $h_B$ with a mass of 400 GeV and $\theta_B\approx10^{-2}$ there will be $\mathcal{O}(1)$ events generated in the high luminosity run at the LHC. 

\begin{figure}[b]
\centering
\includegraphics[width=0.95\linewidth]{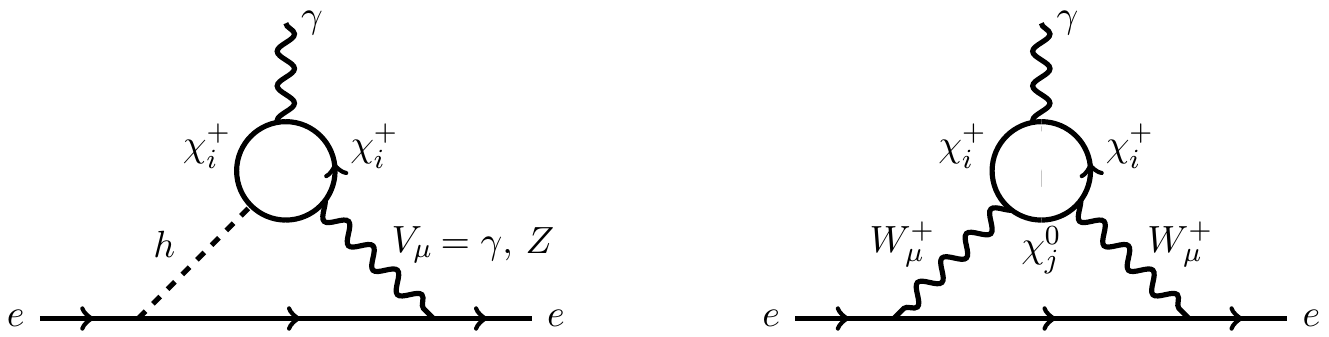}
\caption{Feynman graphs for the Barr-Zee contribution to the electric dipole moment of the electron. $V_\mu$ can either be the photon or the $Z$ boson. Here $\chi_i^+$ and $\chi^0_j$ correspond to the anomaly-canceling fermions.}
\label{fig:DiagEDM}
\end{figure}

\section{CP Violation and EDMs}
\label{sec:EDM}
This theory predicts new fermions needed for anomaly cancellation and their interactions can provide new sources of CP violation. 
In this section we study the new sources of CP violation and calculate the predictions for the electric dipole moment of the electron. These also induce an electric dipole moment for the neutron; however, the experimental bound is much weaker than the one for the electron. In general, all the Yukawa couplings in Eq.~\eqref{eq:Yukawas} can be complex, and as we discuss in Appendix~\ref{sec:appCP} this leads to two independent CP-violating phases: $\phi_C={\rm arg}(y_\eta y_\Psi^* y_1 y_2^*)$ in the charged sector and $\phi_N={\rm arg}(y_\chi y_\Psi^* y_3 y_4^*)$ in the neutral sector. These CP-violating phases contribute to the electric dipole moments of SM fermions via two-loop Barr-Zee diagrams shown in Fig.~\ref{fig:DiagEDM}. The contributions with $h\gamma$, $hZ$ and $hZ_B$ in the loop depend only on $\phi_C$; while both CP-violating phases contribute in the case with $W^-W^+$ in the loop. In Appendix~\ref{sec:EDMcalc} we provide the expressions for the different contributions to the EDM.

\begin{figure}[t]
\centering
\includegraphics[width=0.495\linewidth]{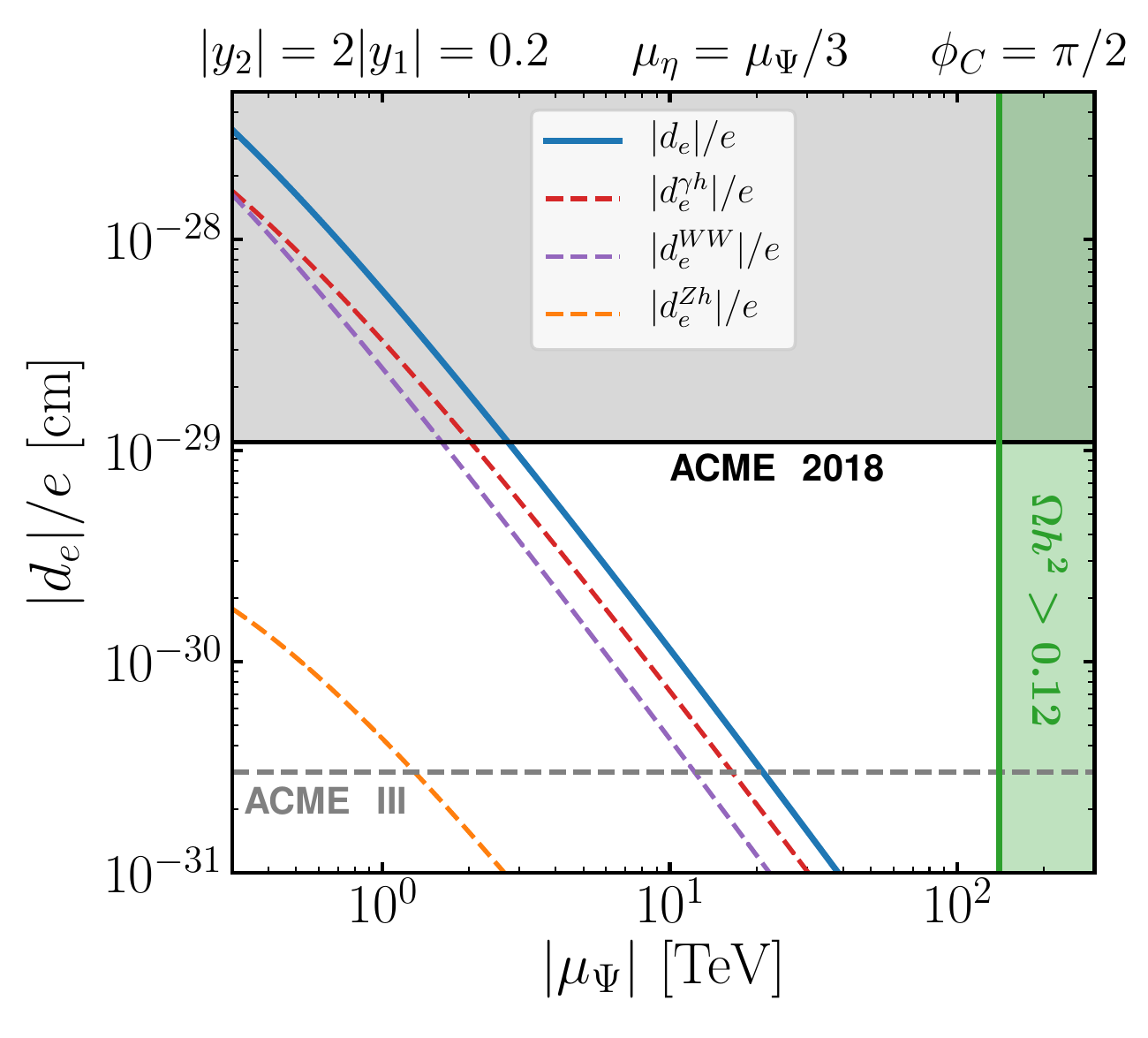}
\includegraphics[width=0.495\linewidth]{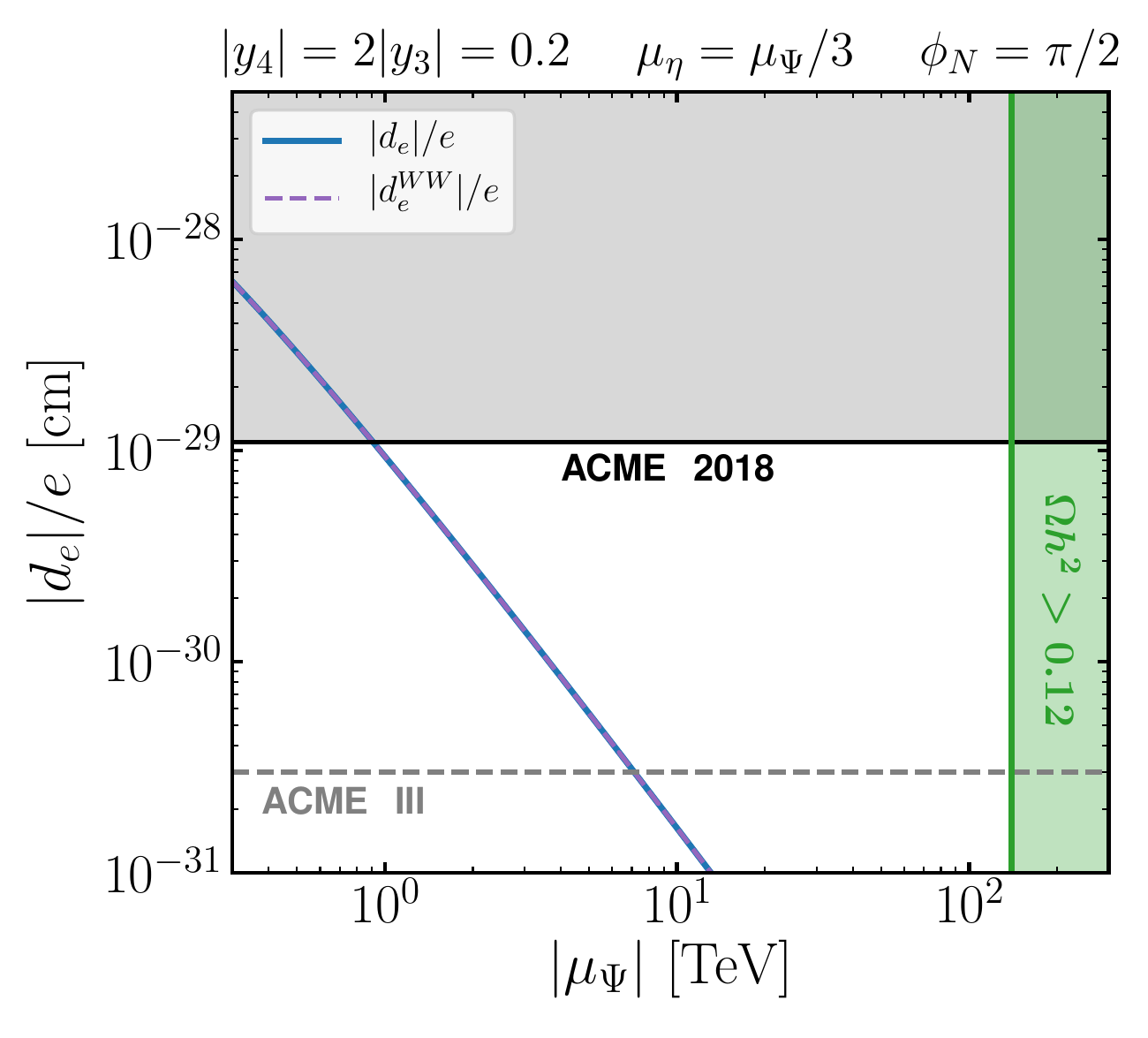}
\caption{\textit{Left panel:} The solid blue line shows the electron EDM as a function of the mass parameter $\mu_\Psi$. The dashed red, purple and orange lines show the different contributions $d_e^{\gamma h}$, $d_e^{WW}$ and $d_e^{Zh}$, respectively. The region shaded in gray corresponds to the experimental bound from the ACME Collaboration~\cite{Andreev:2018ayy} and the gray dashed line gives the projected sensitivity for ACME III~\cite{ACMEIII}. The region shaded in green is excluded due to overproducing the dark matter relic abundance $\Omega_{\rm DM}h^2 > 0.12$. The CP-violating phases are set to $\phi_C=\pi/2$ and $\phi_N=0$ and the other parameters have been fixed as shown in the title of the plot. \textit{Right panel:} Same as left panel but with the CP-violating phases set to $\phi_C=0$ and $\phi_N=\pi/2$.}
\label{fig:eEDM_1D}
\end{figure}

In Fig.~\ref{fig:eEDM_1D} we present our results for the electron EDM as a function of the mass parameter $\mu_\Psi$. The different contributions are shown by dashed lines of different colors. The largest contribution comes from the $\gamma h$ diagram but the $WW $contribution can be of similar order and even larger in some regions of the parameter space. The $Z h$ is small due to the cancellation in the overall term $(T_3^e/2-s_W^2Q_e)$, as has been noted previously in the literature for other scenarios~\cite{Arkani-Hamed:2004zhs}. In the left panel we fix $\phi_C=\pi/2$ and $\phi_N=0$ and the relevant Yukawa interactions to $|y_2|=2|y_1|=0.2$; the region excluded by the ACME collaboration, $|d_e| / e \leq 1.1 \times 10^{-29} \, {\rm cm}$~\cite{Andreev:2018ayy}, corresponds to the region shaded in gray and for this scenario we obtain the exclusion of $|\mu_\Psi| \lesssim 2.7$ TeV. The gray dashed line corresponds to the projected sensitivity for ACME III of $|d_e|/e \leq 3\times 10^{-31}$ cm~\cite{ACMEIII} and for this scenario it will reach $|\mu_\Psi| \lesssim 21$ TeV. The region shaded in green is excluded due to overproducing the dark matter relic abundance $\Omega_{\rm DM}h^2 > 0.12$.

In the right panel of Fig.~\ref{fig:eEDM_1D} we set $\phi_C=0$ and $\phi_N=\pi/2$ which implies that the dominant contribution is from $WW$ in the loop and that is why we can see that the dashed and the solid line overlap. The relevant Yukawa couplings are fixed to $|y_4|=2|y_2|=0.2$; for this scenario we obtain the exclusion of $|\mu_\Psi| \lesssim 0.9$ TeV while ACME III will be able to reach $|\mu_\Psi| \lesssim 7$ TeV. In both panels we fix  $\mu_\eta=\mu_\Psi/3$ and we find that the results are almost independent of $\mu_\chi$ since the $\chi$ fermion is a SM singlet and only gives a small contribution through fermionic mixing. We note that in the limit in which either one of the relevant couplings $y_i\to 0$ then the EDM will also vanish.

\begin{figure}[t]
\centering
\includegraphics[width=0.55\linewidth]{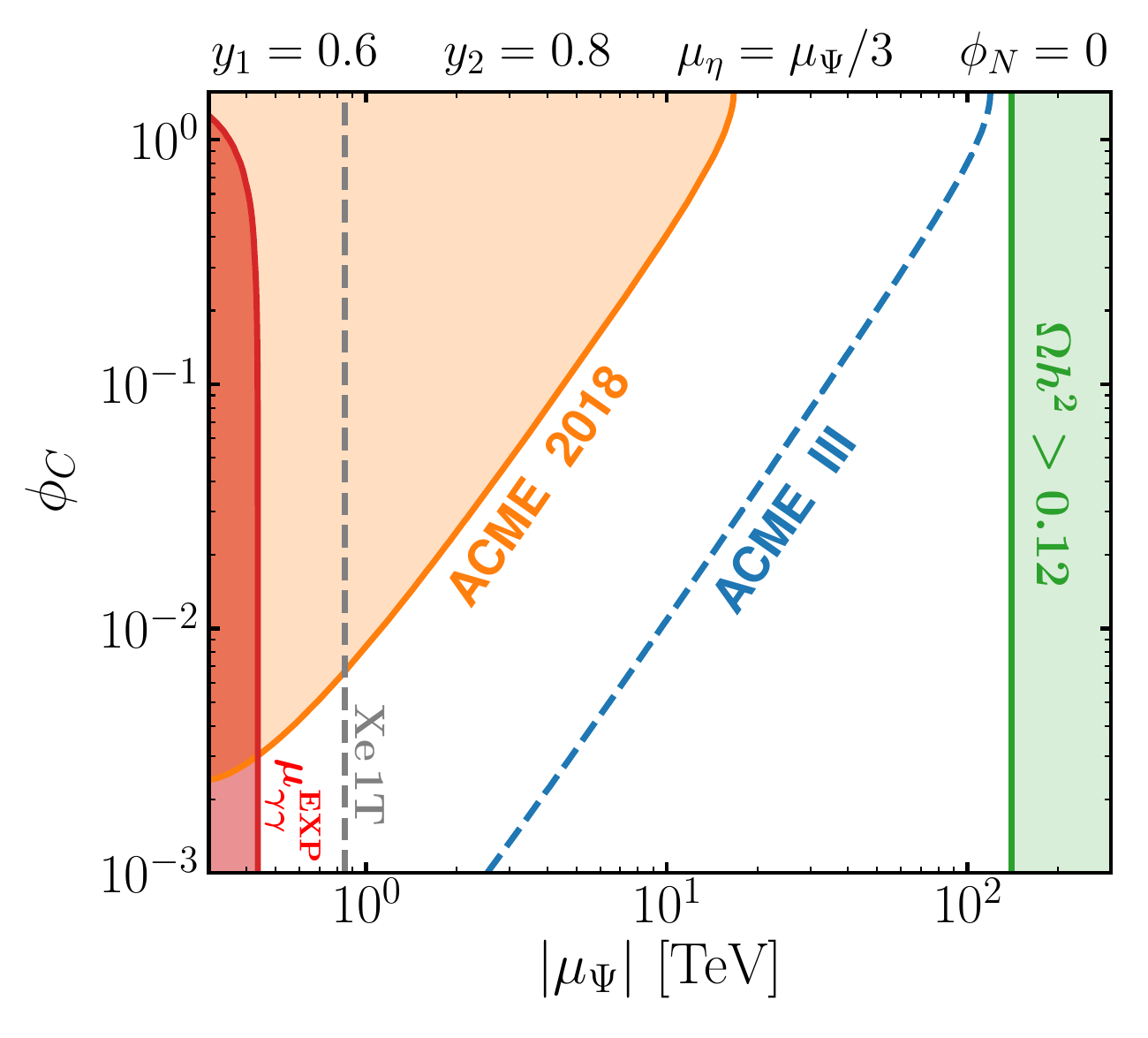}
\caption{ Allowed parameter space in the $\phi_C$ vs $|\mu_\Psi|$ plane. The region shaded in orange corresponds to the experimental bound from the ACME collaboration~\cite{Andreev:2018ayy} and the blue dashed line gives the projected sensitivity for ACME III~\cite{ACMEIII}. The region shaded in red is excluded by the measurement of the Higgs diphoton signal strength $\mu_{\gamma \gamma}$ at the LHC. The gray dashed line gives the bound from DM direct detection, which applies when the dark matter relic density is saturated. The region shaded in green is excluded since it overproduces the dark matter relic abundance $\Omega_{\rm DM} h^2 >0.12$. We set $\phi_N=0$ and the other parameters have been fixed as shown in the title of the plot. }
\label{fig:both}
\end{figure}

These results are relevant for the baryogenesis scenarios proposed in Refs.~\cite{Carena:2018cjh,Carena:2019xrr} where it was argued that EDMs only appear at four loops. We demonstrated that the effects from the anomaly-canceling charged fermions cannot be neglected and that EDMs already appear at two-loops. Nonetheless, our conclusions hold for the minimal scenario and could change in a non-minimal scenario with more scalars charged under $\U(1)_B$ that acquire a non-zero vev. As an alternative solution to the baryon asymmetry of the Universe, Ref.~\cite{Perez:2021udy} recently studied the implementation of high-scale leptogenesis in theories with local lepton and baryon number.
As a side note, we comment on the EDM of the muon which even though it receives and enhancement of ($m_\mu/m_e$) compared to the electron, the experimental sensitivity is much weaker than in the case of the electron. The current bound from Brookhaven Muon ($g-2$) experiment is $|d_\mu|<1.5 \times 10^{-19} \,e \, {\rm cm}$~\cite{Bennett:2008dy} which lies more than six orders of magnitude below our predictions.

In Fig.~\ref{fig:both} we present our results for the parameter space allowed by the ACME bound and the measurement of $\mu_{\gamma \gamma}$. The region excluded by the measurement of the Higgs diphoton channel is shown in red. The gray dashed line is the bound from DM direct detection that requires $M_\chi \gtrsim 850$ GeV; nevertheless, this bound applies when the relic density is saturated. The region excluded by ACME is shown in orange. In the case that $\phi_C=\pi/2$ then this bound already excludes $|\mu_\Psi| \lesssim 16$ TeV. This bound also has consequences for the low-mass regime. Assuming these values for the Yukawas, i.e. $|y_1|=0.6$, $|y_2|=0.8$ and a maximal CP-violating phase then we can find an indirect bound on the symmetry breaking scale by requiring perturbativity of $y_\Psi$ and $y_\eta$; namely, the bound translates into $M_{Z_B}/g_B \gtrsim 13.5$ TeV, so for example for a light boson of $M_{Z_B}=1$ GeV the coupling needs to be $g_B \lesssim 10^{-4}$. The blue dashed line corresponds to the projected sensitivity for ACME III and shows that for maximal CP violation it is expected to reach $|\mu_\Psi| \lesssim 115$ TeV. The region shaded in green overproduces the dark matter relic density. Consequently, the allowed parameter space is constrained from above and below.

This result shows a complementarity between the electron EDM and the Higgs diphoton decay. The region with vanishing CP violation can be probed by measuring a deviation in the Higgs diphoton decay. For the parameters shown in Fig.~\ref{fig:both} the current measurement of $\mu_{\gamma\gamma}$ within $2\sigma$ excludes $|\mu_\Psi| \lesssim 0.37$ TeV. We also find that the projected sensitivity of ACME III will probe a CP-violating phase as small as $\phi_C \approx 10^{-4}$ for charged fermions below the TeV scale.

\section{Summary}
\label{sec:summary}
%
We studied a simple theory based on local baryon number $\U(1)_B$ that predicts a dark matter candidate from the cancellation of gauge anomalies. We found that the constraint on the DM relic density gives an upper bound on the full theory. 
This gauge theory contains new sources of CP violation and can give a large electric dipole moment for the electron. Also, since the new charged fermions cannot be decoupled from the Baryonic Higgs; the theory predicts a large branching ratio for the latter into SM gauge bosons.

Generically, dark matter is predicted to be a Dirac fermion. We studied the dark matter phenomenology and found that the currents bounds from direct detection already require the dark matter mass to be above the TeV scale. We showed that not overproducing the relic density sets a generic upper bound on the masses of the gauge boson and the dark matter, $M_{Z_B}\lesssim 19$ TeV and $M_\chi \lesssim 26$ TeV, respectively. Moreover, since all the new fermions acquire their masses from the new symmetry breaking scale this implies an upper bound on the charged fermions responsible for the EDMs; namely, $M_{\chi_i^\pm} \lesssim 30$ TeV.

We calculated the contribution from the new fermions running in the loop for the decay of the SM Higgs into two photons and found the constraint on the parameter space from the current measurement of the Higgs signal strength $\mu_{\gamma\gamma}$ at the LHC. We also studied the decays for the new Higgs in the theory responsible for the spontaneous breaking of the $\U(1)_B$. We found that the loop-induced decay $h_B \to \gamma \gamma$ can have a large branching ratio around $10\%$ and that the process $p p \to Z_B^* \to Z_B h_B$ could be observed in the high luminosity run at the LHC.

Furthermore, we showed that the theory contains two new CP-violating phases that cannot be removed by field rotations. We computed the Barr-Zee contributions to the EDM of the electron and found that the current bound by ACME already excludes $\mu_\Psi \lesssim 16$ TeV for maximal CP violation. Furthermore, the projected sensitivity for ACME III will be able to probe a large region in the parameter space of the theory; namely, for maximal CP violation it could reach $\mu_\Psi \lesssim 115$ TeV. Consequently, the theory is being constrained from above by the measured dark matter relic abundance and from below by the lack of experimental evidence for an electron EDM. 
Finally, we demonstrated there is a complementarity between measurements of the Higgs diphoton decay and the electron EDM. Namely, the bound from the Higgs diphoton decay is the dominant constraint for small CP-violating phases. This theory could be tested at current or future experiments by combining the results from dark matter, collider and EDM experiments. 

\acknowledgments
A.D.P. is  supported in part by the INFN ``Iniziativa Specifica''   Theoretical Astroparticle Physics  (TAsP-LNF).

\newpage
\appendix
\section{Feynman Rules}
\label{sec:appFR}
In this Appendix we list all the Feynman rules for the new particles:
\beq
\includegraphics[scale=0.65]{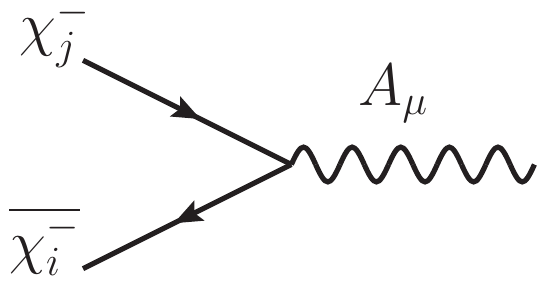} \nonumber 
\eeq
\beq
\hspace{1cm} ie\delta_{ij}\gamma^\mu  \nonumber
\eeq
\\
\beq
\includegraphics[scale=0.65]{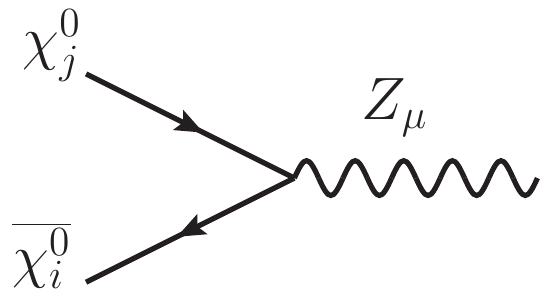} \hspace{3cm} 
\includegraphics[scale=0.65]{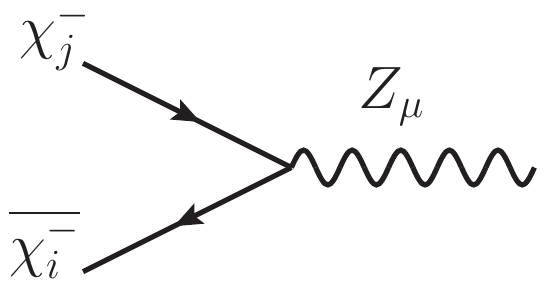} \nonumber 
\eeq
\beq
-\frac{ig \gamma^\mu}{2\cos \theta_W} \left[ (N_L^{2i})^* N_L^{2j} P_L + (N_R^{2i})^* N_R^{2j} P_R  \right]  \hspace{1cm} -i g \sin \theta_W \gamma^\mu \left( C_L^{ij} P_L +  C_R^{ij} P_R \right) \nonumber
\eeq
\\[2ex]
\beq
\includegraphics[scale=0.65]{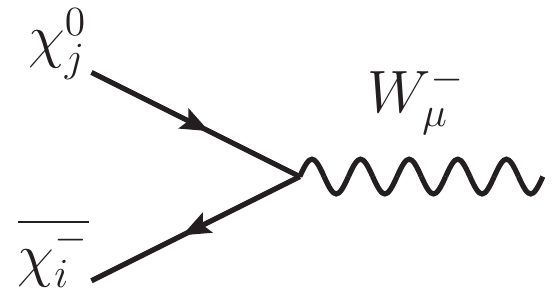} \hspace{3cm} \includegraphics[scale=0.65]{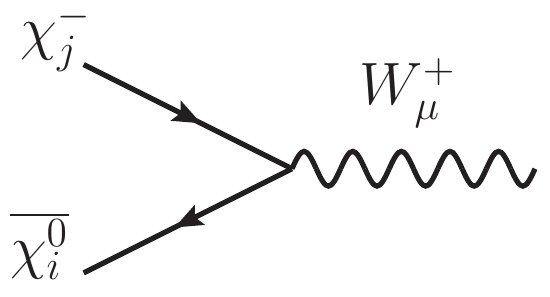} \nonumber 
\eeq
\beq
-\frac{ig \gamma^\mu}{\sqrt{2}} \left[ (V_L^{2i})^* N_L^{2j} P_L + (V_R^{2i})^* N_R^{2j} P_R  \right]   \hspace{1cm} -\frac{ig \gamma^\mu}{\sqrt{2}} \left[  (N_L^{2j})^* V_L^{2j} P_L +  (N_R^{2i})^* V_R^{2j} P_R  \right] \nonumber
\eeq
\\[2ex]
\beq
\includegraphics[scale=0.65]{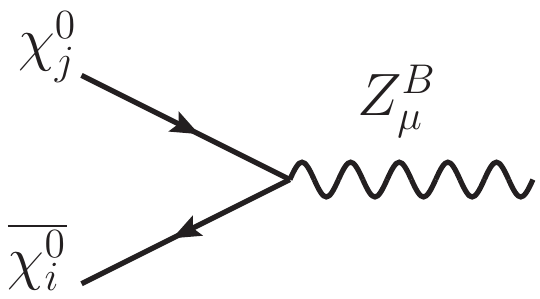} \hspace{3cm} 
\includegraphics[scale=0.65]{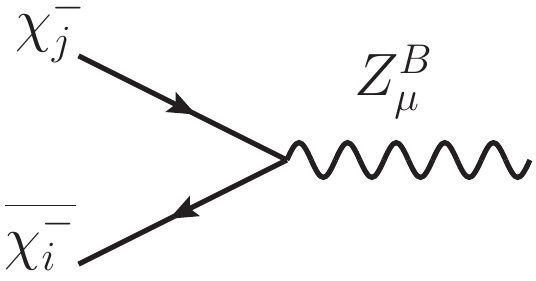} \nonumber 
\eeq

\beq
-i g_B \gamma^\mu \left( O_{Ln}^{ij} P_L +  O_{Rn}^{ij} P_R  \right)   \hspace{1cm} -i g_B \gamma^\mu \left( O_{Lc}^{ij} P_L +  O_{Rc}^{ij} P_R  \right)   \nonumber
\eeq
\\[2ex]
\beq
\includegraphics[scale=0.65]{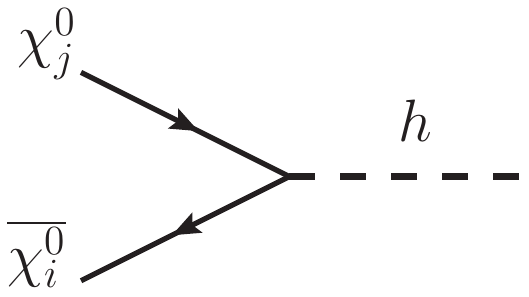} \hspace{3cm} \includegraphics[scale=0.65]{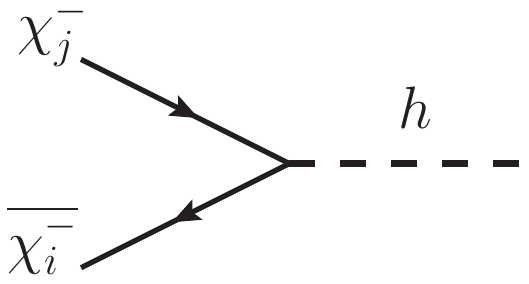} \nonumber 
\eeq
\beq
- i \left[ C^{ij}_{hn} P_L + (C^{ij}_{hn})^* P_R \right] \hspace{3cm} - i \left[ C^{ij}_{hc} P_L + (C^{ij}_{hc})^* P_R \right]\nonumber
\eeq
\\ 
\beq
\includegraphics[scale=0.65]{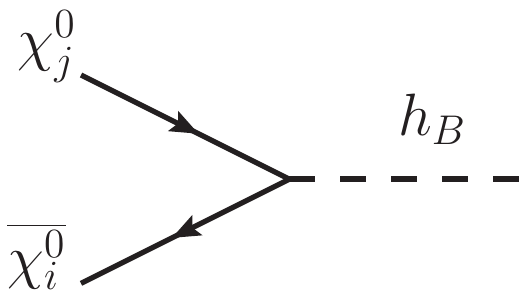} \hspace{3cm} \includegraphics[scale=0.65]{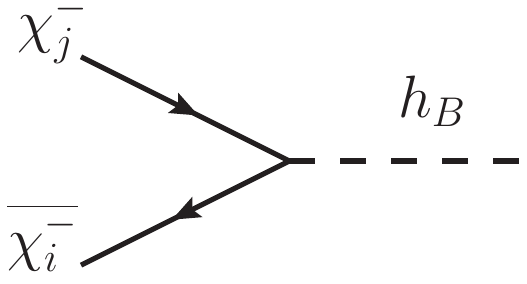} \nonumber 
\eeq
\beq
- i \left[ C^{ij}_{Bn} P_L + (C^{ij}_{Bn})^* P_R \right] \hspace{3cm} - i \left[ C^{ij}_{Bc} P_L + (C^{ij}_{Bc})^* P_R \right]\nonumber
\eeq
\\ \newline where
\begin{align}
C_L^{ij} =& \tan \theta_W (V_L^{1i})^* V_L^{1j} - \frac{1}{\tan 2 \theta_W} (V_L^{2i})^* V_L^{2 j},
\\[1.5ex] 
C_R^{ij} =& \tan \theta_W (V_R^{1i})^* V_R^{1j} - \frac{1}{\tan 2 \theta_W} (V_R^{2i})^* V_R^{2 j} ,
\\[1.5ex]
C^{ij}_{hn} =& \frac{1}{\sqrt{2}} \cos \theta_B \left[ y_3^*  (N_R^{1i})^* N_L^{2j} + y_4 (N_R^{2i})^* N_L^{1j} \right] \nonumber \\
\phantom{=}& -\frac{1}{\sqrt{2}} \sin \theta_B \left[ y_\Psi^*  (N_R^{2i})^* N_L^{2j} + y_\chi (N_R^{1i})^* N_L^{1j} \right],
\\[1.5ex]
C^{ij}_{hc} =& \frac{1}{\sqrt{2}} \cos \theta_B \left[ y_1^* (V_R^{1i})^* V_L^{2j} + y_2 (V_R^{2i})^* V_L^{1j} \right] \nonumber \\
\phantom{=}& -\frac{1}{\sqrt{2}} \sin \theta_B \left[ y_\Psi^* (V_R^{2i})^* V_L^{2j} + y_\eta (V_R^{1i})^* V_L^{1j} \right],
\\[1.5ex]
C^{ij}_{Bn} =& \frac{1}{\sqrt{2}} \sin \theta_B \left[ y_3^*  (N_R^{1i})^* N_L^{2j} + y_4 (N_R^{2i})^* N_L^{1j} \right] \nonumber \\
\phantom{=}& +\frac{1}{\sqrt{2}} \cos \theta_B \left[ y_\Psi^*  (N_R^{2i})^* N_L^{2j} + y_\chi (N_R^{1i})^* N_L^{1j} \right],
\end{align}
\begin{align}
C^{ij}_{Bc} =& \frac{1}{\sqrt{2}} \sin \theta_B \left[ y_1^* (V_R^{1i})^* V_L^{2j} + y_2 (V_R^{2i})^* V_L^{1j} \right] \nonumber \\
\phantom{=}& +\frac{1}{\sqrt{2}} \cos \theta_B \left[ y_\Psi^* (V_R^{2i})^* V_L^{2j} + y_\eta (V_R^{1i})^* V_L^{1j} \right],
\\[1.5ex]
O^{ij}_{Ln} = & B_2  (N_L^{1i})^* N_L^{1j} + B_1 (N_L^{2i})^* N_L^{2j},
\\[2ex]
O^{ij}_{Rn} = &  B_1  (N_R^{1i})^* N_R^{1j} + B_2 (N_R^{2i})^* N_R^{2j},
\\[2ex]
O^{ij}_{Lc} = & B_2  (V_L^{1i})^* V_L^{1j} + B_1 (V_L^{2i})^* V_L^{2j},
\\[2ex]
O^{ij}_{Rc} = &  B_1  (V_R^{1i})^* V_R^{1j} + B_2 (V_R^{2i})^* V_R^{2j},
\end{align}
where the matrices $N_L$, $N_R$, $V_L$ and $V_R$ are the ones that diagonalize the mass matrices as we discuss in Appendix~\ref{sec:diagonal}.

\section{Mass Matrices}
\label{sec:diagonal}
The anomaly-canceling fermions consist of two Dirac neutral fermions and two charged fermions. In this appendix we go through the diagonalization to the physical states. 

\begin{itemize}

\item Neutral States: 

The mass matrix for the neutral states in the basis $\chi_L^0=(\chi_L^0 \,\,\,\,  \Psi_{L}^0 )$ and $\chi_R^0=(\chi_R^0 \,\,\,\,   \Psi_{R}^0)$ is given by
\beq
-\mathcal{L}\supset
 \begin{pmatrix}
\overline{\chi_R^0}  &  \hspace{0.4cm} \overline{\Psi_{R}^0}
 \end{pmatrix} 
 \mathcal{M}_0
  \begin{pmatrix}
\chi_L^0  \\[2ex]  \Psi_{L}^0
 \end{pmatrix}  + 
\text{h.c.} ,
\eeq
where
\beq
\label{eq:M0}
\mathcal{M}_0 = \frac{1}{\sqrt{2}}\begin{pmatrix}
y_\chi v_B \,\,\, & y_3^* v_0 \\[2ex] 
y_4 v_0 \,\,\, &   y_\Psi^* v_B  \end{pmatrix},
\eeq
where $v_0$ and $v_B$ correspond to the vevs of the SM Higgs and the Baryonic Higgs, respectively. To obtain the physical fields $\chi_i^0$ the mass matrix needs to be diagonalized.
The relation between the fields in the Lagrangian and the physical fields is given by the $N_L$ and $N_R$ mixing matrices
\beq
\begin{pmatrix}
\chi_L^0  \\[2ex]   \Psi_{L}^0
 \end{pmatrix} = N_{L} 
 \begin{pmatrix}
  \chi_{1L}^0 \\[2ex] \chi_{2L}^0   
 \end{pmatrix} ,
 \hspace{2cm}
 \begin{pmatrix}
\chi_R^0 \\[2ex]   \Psi_{R}^0 
 \end{pmatrix} = N_{R} 
 \begin{pmatrix}
  \chi_{1R}^0 \\[2ex] \chi_{2R}^0   
 \end{pmatrix} .
\eeq
The unitary matrices $N_L$ and $N_R$ diagonalize the mass matrix as follows
\beq
N_R^\dagger \mathcal{M}_0 N_L = \mathcal{M}_0^{\rm diag},
\eeq
and the following relations can be used to find $N_L$ and $N_R$
\beq
|\mathcal{M}_0^{\rm diag}|^2=N_L^\dagger \mathcal{M}_0^\dagger \mathcal{M}_0 N_L = N_R^\dagger \mathcal{M}_0 \mathcal{M}_0^\dagger N_R,
\eeq
and the two mass eigenvalues correspond to
\begin{align}
M_{\chi^0_i}^2 & = \frac{1}{2} \left[  |\mu_\chi |^2 + |\mu_\Psi |^2 + |y_3|^2 \frac{v_0^2}{2} + |y_4|^2 \frac{v_0^2}{2}  \right. \nonumber \\[1.5ex]
\pm & \left. \sqrt{ \left( |\mu_\chi|^2 + |\mu_\Psi|^2 + |y_3|^2 \frac{v_0^2}{2} + |y_4|^2 \frac{v_0^2}{2} \right)^2 
- 4 |\mu_\chi|^2 |\mu_\Psi|^2 - |y_3|^2 |y_4|^2 v_0^4 + 4 v_0^2 {\rm Re}[\mu_\chi \mu_\Psi^* y_3 y_4^*]  } \, \right] , \nonumber
\end{align}
where $v_0$ corresponds to the SM Higgs vev and we have used $\mu_\chi=y_\chi v_B/\sqrt{2}$ and $\mu_\Psi=y_\Psi v_B/\sqrt{2}$.

\item Charged States: 

The mass matrix for the new charged fermions in the basis $\chi_L^-=(\eta_L^- \,\,\,\,  \Psi_{L}^- )$ and $\chi_R^-=(\eta_R^- \,\,\,\,   \Psi_{R}^-)$ is given by
\beq
-\mathcal{L}\supset
 \begin{pmatrix}
\overline{\eta_R^-}  &  \hspace{0.4cm} \overline{\Psi_{R}^-}
 \end{pmatrix} 
 \mathcal{M}_C
  \begin{pmatrix}
\eta_L^-  \\[2ex]  \Psi_{L}^-
 \end{pmatrix}  + 
\text{h.c.} ,
\eeq
where
\beq
\label{eq:MC}
\mathcal{M}_C = \frac{1}{\sqrt{2}} \begin{pmatrix}
y_\eta v_B\,\,\, & y_1^* v_0 \\[2ex] 
y_2 v_0\,\,\, &  y_\Psi^* v_B 
\end{pmatrix}.
\eeq
To obtain the physical fields $\chi_i^\pm$ the mass matrix needs to be diagonalized.
The relation between the fields in the Lagrangian and the physical fields is given by the $V_L$ and $V_R$ mixing matrices
\beq
\begin{pmatrix}
\eta_L^-  \\[2ex]   \Psi_{L}^-
 \end{pmatrix} = V_{L} 
 \begin{pmatrix}
  \chi_{1L}^- \\[2ex] \chi_{2L}^-   
 \end{pmatrix} ,
 \hspace{2cm}
 \begin{pmatrix}
\eta_R^- \\[2ex]   \Psi_{R}^- 
 \end{pmatrix} = V_{R} 
 \begin{pmatrix}
  \chi_{1R}^- \\[2ex] \chi_{2R}^-   
 \end{pmatrix} .
\eeq
The unitary matrices $V_L$ and $V_R$ diagonalize the mass matrix as follows
\beq
V_R^\dagger \mathcal{M}_C V_L = \mathcal{M}_C^{\rm diag},
\eeq
and the following relations can be used to find $V_L$ and $V_R$
\beq
|\mathcal{M}_C^{\rm diag}|^2=V_L^\dagger \mathcal{M}_C^\dagger \mathcal{M}_C V_L = V_R^\dagger \mathcal{M}_C \mathcal{M}_C^\dagger V_R,
\eeq
and the two mass eigenvalues correspond to
\begin{align}
M_{\chi^\pm_i}^2 & = \frac{1}{2} \left[  |\mu_\eta |^2 + |\mu_\Psi |^2 + |y_1|^2 \frac{v_0^2}{2} + |y_2|^2 \frac{v_0^2}{2}  \right. \nonumber \\[1.5ex]
\pm & \left. \sqrt{ \left( |\mu_\chi|^2 + |\mu_\Psi|^2 + |y_1|^2 \frac{v_0^2}{2} + |y_2|^2 \frac{v_0^2}{2} \right)^2 
- 4 |\mu_\eta|^2 |\mu_\Psi|^2 - |y_1|^2 |y_2|^2 v_0^4 + 4 v_0^2 {\rm Re}[\mu_\eta \mu_\Psi^* y_1 y_2^*]  } \, \right] , \nonumber
\end{align}
where we have used $\mu_\eta=y_\eta v_B/\sqrt{2}$ and $\mu_\Psi=y_\Psi v_B/\sqrt{2}$.

\end{itemize}

\section{CP Violation}
\label{sec:appCP}

The mass matrix for the charged fermions in Eq.~\eqref{eq:MC} can be written in general as 
\beq
\mathcal{M}_C = \begin{pmatrix}
  \displaystyle \frac{|y_\eta| v_B}{\sqrt{2}} e^{i\delta_\eta} \,\,\, & \displaystyle \frac{|y_1| v_0}{\sqrt{2}} e^{-i\delta_1}\\[4ex] 
   \displaystyle \frac{|y_2| v_0}{\sqrt{2}}  e^{i \delta_2}\,\,\, &  \displaystyle \frac{|y_\Psi| v_B}{\sqrt{2}} e^{-i \delta_\Psi} \end{pmatrix},
\eeq
which can be rotated into 
\beq
\mathcal{M}'_C = \begin{pmatrix}
  \displaystyle \frac{|y_\eta| v_B}{\sqrt{2}}  \,\,\, & \displaystyle \frac{|y_1| v_0}{\sqrt{2}} \\[4ex] 
   \displaystyle \frac{|y_2| v_0}{\sqrt{2}}  \,\,\, &  \displaystyle \frac{|y_\Psi| v_B}{\sqrt{2}} e^{i \phi_C} \end{pmatrix},
\eeq
where 
\beq
\phi_C= \delta_\eta - \delta_\Psi + \delta_1 - \delta_2 = {\rm arg}(y_\eta y_\Psi^* y_1 y_2^*),
\eeq
and the result of the EDMs are proportional to this CP-violating phase. A similar rotation can be performed to isolate $\phi_C$ to any matrix entry and this phase cannot be removed by phase rotations. Applying a similar procedure to the mass matrix for the neutral states, given in Eq.~\eqref{eq:M0}, we conclude that there is another CP-violating phase in the neutral sector given by \beq
\phi_N=\delta_\chi - \delta_\Psi + \delta_3 - \delta_4 = {\rm arg}(y_\chi y_\Psi^* y_3 y_4^*).
\eeq

\section{Contributions to the EDMs}
\label{sec:EDMcalc}
For the two-loop calculation of the diagram in Fig.~\ref{fig:DiagEDM} we follow the approach presented in Ref.~\cite{Nakai:2016atk}. The different contributions to the EDM of a SM fermion $f$ are given by
\begin{align}
d_f^{\gamma h} & =  \frac{\alpha^2 \cos \theta_B Q_f }{4 \pi^2 s_W } \frac{m_f}{m_h^2 \, m_W} \sum_{i=1}^2 M_{\chi_i^\pm} {\rm Im}[C_h^{ii}] \, I_{\gamma h}^i(M_{\chi_i^\pm}) ,
\\[1.5ex]
d_f^{hZ} & =  \frac{e \alpha \cos \theta_B}{32 \pi^2 c_W s_W^2} \left( \frac{1}{2}T_3^f - s_W^2 Q_f \right) \frac{m_f}{m_h^2 m_W} \sum_{i,j=1}^2  I_{hZ}^{ij}(M_{\chi_i^\pm}, M_{\chi_j^\pm}), \\[1.5ex]
d_f^{W W} & = \frac{T_3^f \alpha^2 e}{8 \pi^2 s_W^4} \sum_{i,j=1}^2 {\rm Im[} (V_L^{2i})^* N_L^{2j} V_R^{2i} (N_R^{2j})^* ] \frac{m_f M_{\chi^\pm_i} M_{\chi^0_j} }{m_W^4} \, I_{WW}^{ij}(M_{\chi_i^\pm}, M_{\chi_j^0}), \\[1.5ex]
d_f^{hZ_B} & = \frac{ g_B
 \alpha \cos \theta_B}{96 \pi^2 s_W} \frac{m_f}{m_h^2 m_W} \sum_{i,j=1}^2  I_{hZ_B}^{ij}(M_{\chi_i^\pm}, M_{\chi_j^\pm}),
\end{align}
where $M_{\chi_i^\pm}$ correspond to the physical masses of the anomaly-canceling fermions and the loop integrals $I(m)$ are given by
\begin{align}
I_{\gamma h}^i(M_{\chi_i^\pm}) = & \int_0^1  \frac{dx}{x} j\left(0, \frac{M^2_{\chi_i^\pm}}{m_h^2} \frac{1}{x(1-x)} \right), \nonumber
\end{align}
\begin{align}
I_{hZ}^{ij}(M_{\chi_i^\pm}, M_{\chi_j^\pm}) = & \int_0^1 dx \frac{1}{x(1-x)}  j\left(\frac{m_Z^2}{m_h^2
}, \frac{\Delta_{ij}(x)}{m_h^2} \right) \nonumber \\[1ex]
 {\rm Re} \left[(i g_S^{ji} \right. &  \left. (g_A^{ij})^* - g_P^{ji} \, (g_V^{ij})^*) M_{\chi^\pm_i} x(1-x)   -  (i g_S^{ji} \, (g_A^{ij})^* + g_P^{ji} \, (g_V^{ij})^*) M_{\chi^\pm_j} (1-x)^2 \right], \nonumber \\[1.5ex]
I_{WW}^{ij}(M_{\chi_i^\pm}, M_{\chi_j^0}) = & \int_0^1 \frac{dx}{1-x} j\left(0, \frac{xr_{\chi^\pm_i} + (1-x) r_{\chi^0_j}}{x(1-x)} \right),\nonumber \\[1ex]
I_{hZ_B}^{ij}(M_{\chi_i^\pm}, M_{\chi_j^\pm}) = & \int_0^1 dx \frac{1}{x(1-x)}  j\left(\frac{M_{Z_B}^2}{m_h^2
}, \frac{\Delta_{ij}(x)}{m_h^2} \right) \nonumber \\[1ex]
{\rm Re} \left[(i g_S^{ji} \right. & \left.  (\kappa_A^{ij})^* - g_P^{ji} \, (\kappa_V^{ij})^*) M_{\chi^\pm_i} x(1-x)   -  (i g_S^{ji} \, (\kappa_A^{ij})^* + g_P^{ji} \, (\kappa_V^{ij})^*) M_{\chi^\pm_j} (1-x)^2 \right], \nonumber 
\end{align}
with the parameters
\begin{align}
g_S^{ij}  =  - {\rm Re} [ C_h^{ij} ],
& \hspace{1cm}
g_P^{ij}  = \,\, {\rm Im} [ C_h^{ij} ],
\\[1.5ex]
g_V^{ij}  =  - \frac{g \sin \theta_W}{2} \left( C_L^{ij} + C_R^{ij} \right),
& \hspace{1cm}
g_A^{ij} =  \,\, \frac{g \sin \theta_W}{2} \left( C_L^{ij} - C_R^{ij}  \right),
\\[1.5ex]
\kappa_V^{ij}  = -\frac{g_B}{2} \left( O_L^{ij} + O_R^{ij} \right),
& \hspace{1cm}
\kappa_A^{ij} = -\frac{g_B}{2} \left( O_R^{ij} - O_L^{ij}  \right),
\end{align}
where the matrices $C_h$, $C_L$, $C_R$, $O_L$ and $O_R$ are given in Appendix~\ref{sec:appFR}. $\theta_W$ corresponds to the Weinberg angle and 
the functions in the integrands correspond to 
\begin{align}
j(y,z)= & \frac{1}{y-z} \left( \frac{y\log y}{y-1} - \frac{z\log z}{z-1}  \right), \\[1.5ex]
\Delta_{ij}(x) = &  \frac{x M_{\chi_i^\pm}^2 + (1-x) M_{\chi_j^\pm}^2}{x(1-x)}, \\[1.5ex]
r_{\chi_i}=&\left( \frac{M_{\chi_i}}{m_W} \right)^2.
\end{align}
%

\section{Baryonic Higgs}
\label{sec:appDecaysHB}
In this Appendix, we present the expressions for the decays of the Baryonic Higgs including the loop-induced channels and the full analytic expressions. To compute the loop functions we make use of the Package-X~\cite{Patel:2015tea} Mathematica package. In the Feynman diagrams below, we use an orange dot for vertices that can only occur by the mixing of the Baryonic Higgs with the SM Higgs.
\begin{itemize}[leftmargin=3mm]
%
\item ${\boldmath h_B \to \gamma \gamma}$:
\begin{eqnarray}
\imineq{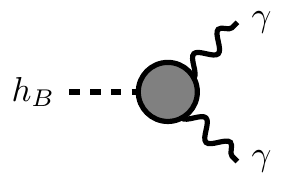}{11} &=& \imineq{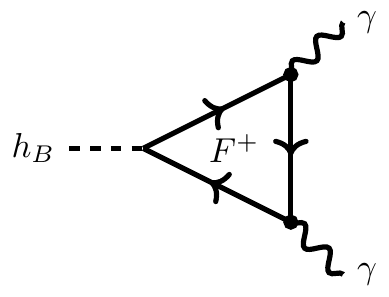}{17}+\imineq{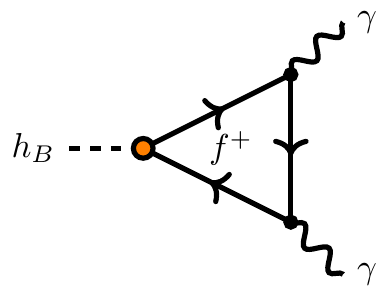}{17} + \imineq{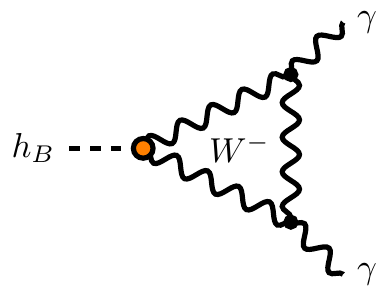}{17} \nonumber  \\
\Gamma(h_B \to \gamma \gamma) &=&  \frac{\alpha^2}{64  \pi^3  M_{h_B}^5} \left |  \cos \theta_B \sum_{F^+}  \frac{3 g_B M_{F^+}^2}{M_{Z_B}} F_{F^+} + \frac{\sin \theta_B}{v_0} \left( \sum_{f^+} N_c^f Q_f^2 m_{f^+}^2 F_{f^+} -   F_W \right) \right |^2.  \quad \quad \quad \nonumber 
\end{eqnarray}
%
\item  $h_B \to Z \gamma$:
\begin{eqnarray}
 \!\!\!\!  \imineq{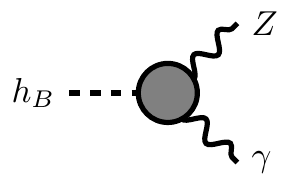}{11} \!\!\!\!\! &=& \imineq{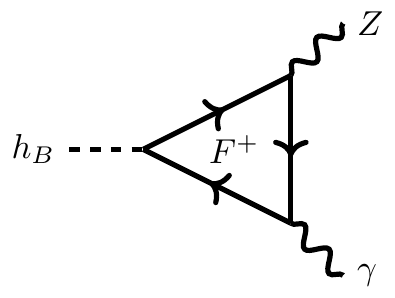}{17} \!\!\!\!\! + \imineq{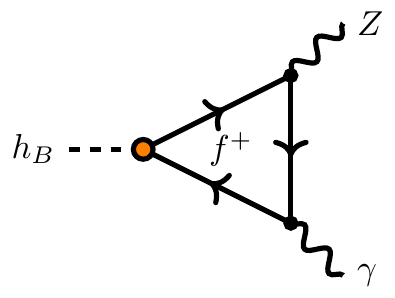}{17} \!\!\!\!\!+  \imineq{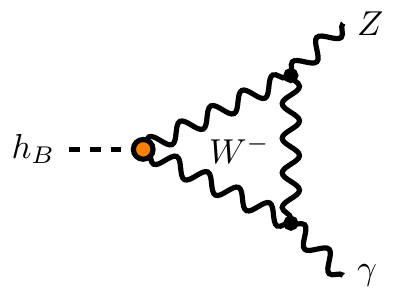}{17} \quad \quad \quad \quad \quad \quad  \quad  \quad \quad \quad \quad   \nonumber \\
  \!\!\!\!  \!\!\!\! \Gamma(h_B \to Z \gamma ) &=& \frac{\alpha^2}{8\pi^3} \frac{M_{h_B}^2 \! - \! M_{Z}^2}{M_{h_B}^3}  \left |  \frac{\sin \theta_B}{v_0 \, s2\theta_W} \left( \sum_{u}  m_u^2 \left(1 \! - \! \frac{8}{3} \sin^2 \theta_W\right)  A_u + \sum_{d} \frac{m_d^2}{2} \left(1 \! - \! \frac{4}{3}\sin^2 \theta_W \right) A_d   \right. \right. \quad \quad \quad \nonumber \\
& + & \left. \sum_{e} \frac{m_e^2}{2} \left(1 \! - \! 4 \sin^2 \theta_W \right) A_e - \frac{\cos^2\theta_W}{2} A_W \right)
+ \left. \frac{3 \, g_B\cos \theta_B}{M_{Z_B}} \left( t \theta_W M_{\eta^+}^2  A_{\eta^+} +  \frac{M_{\Psi^+}^2}{t2\theta_W}A_{\Psi^+} \right) \right |^2. \nonumber
\end{eqnarray}  
%
\item $h_B \to WW$:
%
\begin{eqnarray}
 \imineq{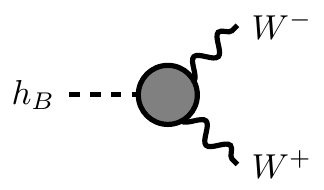}{12} &=& \imineq{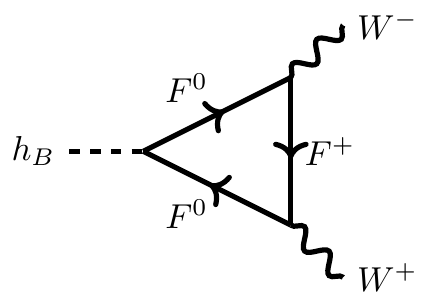}{15}+\imineq{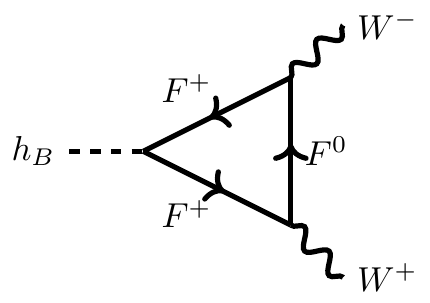}{15}+ \imineq{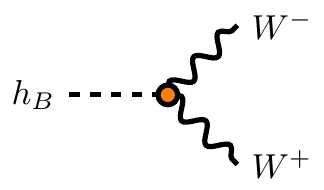}{12} \nonumber \\
\Gamma (h_B \to WW) &=& \frac{\sqrt{M_{h_B}^2 - 4M_W^2}}{16\pi M_{h_B}^2 M_W^4} \left (\cos^2 \theta_B \frac{9g_B^2}{M_{Z_B}^2}\left | \frac{g^2}{2} B_\Psi[W] \right |^2 +2(M_{h_B}^2-2M_W^2) \times \right. \nonumber \\
&&\text{Re} \left \{ \cos \theta_B \frac{3 g_B}{M_{Z_B}} \frac{g^2}{2} B_\Psi[W] \left ( \cos \theta_B \frac{3g_B}{M_{Z_B}} \frac{g^2}{2} C_\Psi^*[W] + \frac{\sin \theta_B}{v_0} M_W^2 \right) \right \}  \nonumber \\
&& \left.+ \left |  \cos \theta_B \frac{3g_B}{M_{Z_B}} \frac{g^2}{2} C_\Psi[W] +  \frac{\sin \theta_B}{v_0} M_W^2 \right|^2(M_{h_B}^4 - 4 M_{h_B}^2 M_W^2 + 12 M_W^4) \right). \nonumber
\end{eqnarray}
\item $h_B \to ZZ$:
%
\begin{eqnarray}
\imineq{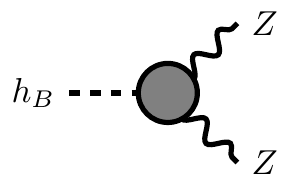}{12}
&=&
\imineq{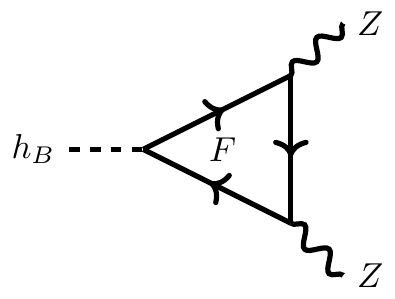}{15}
+
\imineq{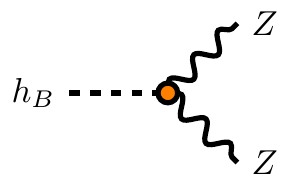}{12} \nonumber
\\
\Gamma (h_B \to ZZ) &=& \frac{\sqrt{M_{h_B}^2 - 4M_Z^2}}{128\pi M_{h_B}^2 M_Z^4} \left ( \cos^2\theta_B \frac{9g_B^2}{M_{Z_B}^2}\left |\sum_F g_{ZF}^2 B_F[Z] \right |^2 + 2 (M_{h_B}^2-2M_Z^2) \times \right. \nonumber  \\
&&\text{Re} \left \{ \cos \theta_B \frac{3 g_B}{M_{Z_B}} \sum_F g_{ZF}^2  B_F[Z] \left ( \cos \theta_B \frac{3g_B}{M_{Z_B}} \sum_F g_{ZF}^2 C_F^*[Z] + 2 \frac{\sin \theta_B}{v_0} M_Z^2 \right) \right \} \nonumber \\
&& \left.+ \left |\cos \theta_B \frac{3g_B}{M_{Z_B}} \sum_F g_{ZF}^2 C_F[Z]  + 2  \frac{\sin \theta_B}{v_0} M_Z^2 \right|^2(M_{h_B}^4 - 4 M_{h_B}^2 M_Z^2 + 12 M_Z^4) \right). \nonumber
\end{eqnarray}
%
\end{itemize}
In all these formulae $F_i$ and $f_i$ correspond to the anomaly-canceling and the SM fermions, respectively.
The couplings between the gauge bosons and the new anomaly-canceling fermions are given by (see Feynman rules in Appendix~\ref{sec:appFR}),
\begin{equation*}
g_{Z\Psi^0} = \frac{e}{\sin 2 \theta_W}, \quad \quad g_{Z\Psi^+} = -\frac{e}{\tan 2 \theta_W}, \quad \quad g_{Z\eta^+} = e \tan \theta_W.
 \end{equation*}

\section{Loop Functions}
\label{sec:appDecays}

In this Appendix we present the loop functions for the decay $h \to \gamma \gamma$ given in Eq.\eqref{eq:haa} and the decays of $h_B$ given in Appendix~\ref{sec:appDecaysHB}. We calculated these functions with the Package-X~\cite{Patel:2015tea} Mathematica package. We write explicitly the loop functions as a function of the fermion mass $M_f$ or gauge boson mass $M_V$ running inside the loop and the Passarino-Veltman functions defined in Package-X:
\begin{itemize}
 \item Loop functions entering in the $h_B$ and $h$ decays to two massless gauge fields:
\begin{eqnarray*}
 F_f &=&  4 M_{h}^2 + (4 M_f^2 - M_{h}^2) \log^2 \left[1-\frac{M_{h}^2}{2M_f^2}\left( 1 - \sqrt{1 - \frac{4M_f^2}{M_{h}^2 \, }} \right) \right],\\[1.5ex]
 F_W &=& M_{h}^4 + 6 M_{h}^2 M_W^2 - 3 (M_{h}^2 - 2 M_W^2)M_W^2 \log^2 \left[1- \frac{M_{h}^2}{2M_W^2}\left( 1 - \sqrt{1 -\frac{ 4M_W^2}{M_{h}^2}} \, \right)\right],\\[1.5ex]
 G_{\chi_i^\pm} &=&  \log^2 \left[1-\frac{M_{h}^2}{2M_{\chi_i^\pm}^2}\left( 1 - \sqrt{1 - \frac{4M_{\chi_i^\pm}^2}{M_{h}^2 \, }} \right) \right].
 \end{eqnarray*}
 \item Loop functions entering in the $h_B$ decay to a massive and a massless gauge fields:
 \begin{eqnarray*}
 A_f &=& (4m_f^2-M_{h_B}^2+M_{V}^2)C_0[0,M_{h_B}^2,M_{V}^2;m_f] + 2\left(\frac{M_{V}^2(\Lambda[M_{h_B}^2]-\Lambda[M_{V}^2])}{M_{h_B}^2- M^2_{V}}+1\right),\\[2.5ex]
 A_W &=& 2 \, M_W^2 \, C_0[0,M_{h_B}^2,M_{V}^2;M_W] \left( (M_{h_B}^2-2M_W^2)(\tan^2\theta_W-5) - 2 M_V^2 (\tan^2\theta_W-3)\right)\\
 && +\left( \frac{M_V^2}{M_{h_B}^2-M_V^2} (\Lambda[M_{h_B}^2]-\Lambda[M_V^2]) +1\right) \left(M_{h_B}^2(1-\tan^2\theta_W)+2M_W^2 (5-\tan^2\theta_W)\right).
 \end{eqnarray*}
 \item Loop functions entering in the $h_B$ decay to a two equal massive gauge bosons:
 \begin{eqnarray*}
 B_F[V] &=&\frac{1}{16 \pi^2} \frac{4M_F^2}{M_{h_B}^2-4M_V^2} \times \\[1.5ex]
 && \left[ (M_{h_B}^2-2M_V^2) \left( M_{h_B}^4 - 4 M_F^2 (M_{h_B}^2-4M_V^2) - 6 M_{h_B}^2M_V^2-4M_V^4\right) C_0[M_{h_B}^2,M_V^2,M_V^2;M_F] \right. \\[1.5ex]
&& \left. + \, 4 M_V^2 (M_{h_B}^2 + 2M_V^2)(\Lambda[M_V^2]-\Lambda[M_{h_B}^2])-2(M_{h_B}^4-6M_{h_B}^2M_V^2+8M_V^4)\right],\\[2.5ex]
C_F[V] &=& \frac{1}{16 \pi^2} \frac{4M_F^2}{M_{h_B}^2-4M_V^2} \left[ 2 (M_{h_B}^2 + 2M_V^2 (\Lambda[M_{h_B}^2]-\Lambda[M_V^2]) -4M_V^2)\right. \\
&& \left. - (M_{h_B}^4 - 6 M_{h_B}^2 M_V^2 - 4 M_F^2 (M_{h_B}^2 - 4 M_V^2) + 4 M_V^4)C_0 [M_{h_B}^2,M_V^2,M_V^2;M_F]\right].
\end{eqnarray*}
\end{itemize}

\bibliographystyle{JHEP}
\bibliography{EDM-6reps}{}

\providecommand{\href}[2]{#2}\begingroup\raggedright\begin{thebibliography}{10}

\bibitem{Duerr:2013dza}
M.~Duerr, P.~Fileviez~Perez and M.~B. Wise, \emph{{Gauge Theory for Baryon and
  Lepton Numbers with Leptoquarks}},
  \href{https://doi.org/10.1103/PhysRevLett.110.231801}{\emph{Phys. Rev. Lett.}
  {\bfseries 110} (2013) 231801},
  [\href{https://arxiv.org/abs/1304.0576}{{\ttfamily 1304.0576}}].

\bibitem{Perez:2014qfa}
P.~Fileviez~Perez, S.~Ohmer and H.~H. Patel, \emph{{Minimal Theory for
  Lepto-Baryons}},
  \href{https://doi.org/10.1016/j.physletb.2014.06.057}{\emph{Phys. Lett.}
  {\bfseries B735} (2014) 283--287},
  [\href{https://arxiv.org/abs/1403.8029}{{\ttfamily 1403.8029}}].

\bibitem{Duerr:2013lka}
M.~Duerr and P.~Fileviez~Perez, \emph{{Baryonic Dark Matter}},
  \href{https://doi.org/10.1016/j.physletb.2014.03.011}{\emph{Phys. Lett. B}
  {\bfseries 732} (2014) 101--104},
  [\href{https://arxiv.org/abs/1309.3970}{{\ttfamily 1309.3970}}].

\bibitem{Duerr:2014wra}
M.~Duerr and P.~Fileviez~Perez, \emph{{Theory for Baryon Number and Dark Matter
  at the LHC}}, \href{https://doi.org/10.1103/PhysRevD.91.095001}{\emph{Phys.
  Rev. D} {\bfseries 91} (2015) 095001},
  [\href{https://arxiv.org/abs/1409.8165}{{\ttfamily 1409.8165}}].

\bibitem{FileviezPerez:2015mlm}
P.~Fileviez~Perez, \emph{{New Paradigm for Baryon and Lepton Number
  Violation}}, \href{https://doi.org/10.1016/j.physrep.2015.09.001}{\emph{Phys.
  Rept.} {\bfseries 597} (2015) 1--30},
  [\href{https://arxiv.org/abs/1501.01886}{{\ttfamily 1501.01886}}].

\bibitem{FileviezPerez:2018jmr}
P.~Fileviez~Perez, E.~Golias, R.-H. Li and C.~Murgui, \emph{{Leptophobic Dark
  Matter and the Baryon Number Violation Scale}},
  \href{https://doi.org/10.1103/PhysRevD.99.035009}{\emph{Phys. Rev. D}
  {\bfseries 99} (2019) 035009},
  [\href{https://arxiv.org/abs/1810.06646}{{\ttfamily 1810.06646}}].

\bibitem{Andreev:2018ayy}
{\scshape ACME} collaboration, V.~Andreev et~al., \emph{{Improved limit on the
  electric dipole moment of the electron}},
  \href{https://doi.org/10.1038/s41586-018-0599-8}{\emph{Nature} {\bfseries
  562} (2018) 355--360}.

\bibitem{Bernreuther:1990jx}
W.~Bernreuther and M.~Suzuki, \emph{{The electric dipole moment of the
  electron}}, \href{https://doi.org/10.1103/RevModPhys.63.313}{\emph{Rev. Mod.
  Phys.} {\bfseries 63} (1991) 313--340}.

\bibitem{Pospelov:2005pr}
M.~Pospelov and A.~Ritz, \emph{{Electric dipole moments as probes of new
  physics}}, \href{https://doi.org/10.1016/j.aop.2005.04.002}{\emph{Annals
  Phys.} {\bfseries 318} (2005) 119--169},
  [\href{https://arxiv.org/abs/hep-ph/0504231}{{\ttfamily hep-ph/0504231}}].

\bibitem{Fukuyama:2012np}
T.~Fukuyama, \emph{{Searching for New Physics beyond the Standard Model in
  Electric Dipole Moment}},
  \href{https://doi.org/10.1142/S0217751X12300153}{\emph{Int. J. Mod. Phys. A}
  {\bfseries 27} (2012) 1230015},
  [\href{https://arxiv.org/abs/1201.4252}{{\ttfamily 1201.4252}}].

\bibitem{Chupp:2017rkp}
T.~Chupp, P.~Fierlinger, M.~Ramsey-Musolf and J.~Singh, \emph{{Electric dipole
  moments of atoms, molecules, nuclei, and particles}},
  \href{https://doi.org/10.1103/RevModPhys.91.015001}{\emph{Rev. Mod. Phys.}
  {\bfseries 91} (2019) 015001},
  [\href{https://arxiv.org/abs/1710.02504}{{\ttfamily 1710.02504}}].

\bibitem{Perez:2020jyg}
P.~Fileviez~Perez and A.~D. Plascencia, \emph{{Electric dipole moments, new
  forces and dark matter}},
  \href{https://doi.org/10.1007/JHEP03(2021)185}{\emph{JHEP} {\bfseries 03}
  (2021) 185}, [\href{https://arxiv.org/abs/2008.09116}{{\ttfamily
  2008.09116}}].

\bibitem{FileviezPerez:2019jju}
P.~Fileviez~Perez, E.~Golias, R.-H. Li, C.~Murgui and A.~D. Plascencia,
  \emph{{Anomaly-free dark matter models}},
  \href{https://doi.org/10.1103/PhysRevD.100.015017}{\emph{Phys. Rev. D}
  {\bfseries 100} (2019) 015017},
  [\href{https://arxiv.org/abs/1904.01017}{{\ttfamily 1904.01017}}].

\bibitem{Planck:2018vyg}
{\scshape Planck} collaboration, N.~Aghanim et~al., \emph{{Planck 2018 results.
  VI. Cosmological parameters}},
  \href{https://doi.org/10.1051/0004-6361/201833910}{\emph{Astron. Astrophys.}
  {\bfseries 641} (2020) A6},
  [\href{https://arxiv.org/abs/1807.06209}{{\ttfamily 1807.06209}}].

\bibitem{ATLAS:2017eqx}
{\scshape ATLAS} collaboration, M.~Aaboud et~al., \emph{{Search for new
  phenomena in dijet events using 37 fb$^{-1}$ of $pp$ collision data collected
  at $\sqrt{s}=$13 TeV with the ATLAS detector}},
  \href{https://doi.org/10.1103/PhysRevD.96.052004}{\emph{Phys. Rev. D}
  {\bfseries 96} (2017) 052004},
  [\href{https://arxiv.org/abs/1703.09127}{{\ttfamily 1703.09127}}].

\bibitem{CMS:2018mgb}
{\scshape CMS} collaboration, A.~M. Sirunyan et~al., \emph{{Search for narrow
  and broad dijet resonances in proton-proton collisions at $ \sqrt{s}=13 $ TeV
  and constraints on dark matter mediators and other new particles}},
  \href{https://doi.org/10.1007/JHEP08(2018)130}{\emph{JHEP} {\bfseries 08}
  (2018) 130}, [\href{https://arxiv.org/abs/1806.00843}{{\ttfamily
  1806.00843}}].

\bibitem{XENON:2018voc}
{\scshape XENON} collaboration, E.~Aprile et~al., \emph{{Dark Matter Search
  Results from a One Ton-Year Exposure of XENON1T}},
  \href{https://doi.org/10.1103/PhysRevLett.121.111302}{\emph{Phys. Rev. Lett.}
  {\bfseries 121} (2018) 111302},
  [\href{https://arxiv.org/abs/1805.12562}{{\ttfamily 1805.12562}}].

\bibitem{XENON:2015gkh}
{\scshape XENON} collaboration, E.~Aprile et~al., \emph{{Physics reach of the
  XENON1T dark matter experiment}},
  \href{https://doi.org/10.1088/1475-7516/2016/04/027}{\emph{JCAP} {\bfseries
  04} (2016) 027}, [\href{https://arxiv.org/abs/1512.07501}{{\ttfamily
  1512.07501}}].

\bibitem{Belanger:2018ccd}
G.~B\'elanger, F.~Boudjema, A.~Goudelis, A.~Pukhov and B.~Zaldivar,
  \emph{{micrOMEGAs5.0 : Freeze-in}},
  \href{https://doi.org/10.1016/j.cpc.2018.04.027}{\emph{Comput. Phys. Commun.}
  {\bfseries 231} (2018) 173--186},
  [\href{https://arxiv.org/abs/1801.03509}{{\ttfamily 1801.03509}}].

\bibitem{ATLAS:2018hxb}
{\scshape ATLAS} collaboration, M.~Aaboud et~al., \emph{{Measurements of Higgs
  boson properties in the diphoton decay channel with 36 fb$^{-1}$ of $pp$
  collision data at $\sqrt{s} = 13$ TeV with the ATLAS detector}},
  \href{https://doi.org/10.1103/PhysRevD.98.052005}{\emph{Phys. Rev. D}
  {\bfseries 98} (2018) 052005},
  [\href{https://arxiv.org/abs/1802.04146}{{\ttfamily 1802.04146}}].

\bibitem{CMS:2021kom}
{\scshape CMS} collaboration, A.~M. Sirunyan et~al., \emph{{Measurements of
  Higgs boson production cross sections and couplings in the diphoton decay
  channel at $ \sqrt{\mathrm{s}} $ = 13 TeV}},
  \href{https://doi.org/10.1007/JHEP07(2021)027}{\emph{JHEP} {\bfseries 07}
  (2021) 027}, [\href{https://arxiv.org/abs/2103.06956}{{\ttfamily
  2103.06956}}].

\bibitem{CMS:2013xfa}
{\scshape CMS} collaboration, \emph{{Projected Performance of an Upgraded CMS
  Detector at the LHC and HL-LHC: Contribution to the Snowmass Process}},  in
  \emph{{Community Summer Study 2013}: {Snowmass on the Mississippi}}, 7, 2013,
  \href{https://arxiv.org/abs/1307.7135}{{\ttfamily 1307.7135}}.

\bibitem{McKeen:2012av}
D.~McKeen, M.~Pospelov and A.~Ritz, \emph{{Modified Higgs branching ratios
  versus CP and lepton flavor violation}},
  \href{https://doi.org/10.1103/PhysRevD.86.113004}{\emph{Phys. Rev. D}
  {\bfseries 86} (2012) 113004},
  [\href{https://arxiv.org/abs/1208.4597}{{\ttfamily 1208.4597}}].

\bibitem{Korchin:2013ifa}
A.~Y. Korchin and V.~A. Kovalchuk, \emph{{Polarization effects in the Higgs
  boson decay to $\gamma Z$ and test of $CP$ and $CPT$ symmetries}},
  \href{https://doi.org/10.1103/PhysRevD.88.036009}{\emph{Phys. Rev. D}
  {\bfseries 88} (2013) 036009},
  [\href{https://arxiv.org/abs/1303.0365}{{\ttfamily 1303.0365}}].

\bibitem{Alloul:2013naa}
A.~Alloul, B.~Fuks and V.~Sanz, \emph{{Phenomenology of the Higgs Effective
  Lagrangian via FEYNRULES}},
  \href{https://doi.org/10.1007/JHEP04(2014)110}{\emph{JHEP} {\bfseries 04}
  (2014) 110}, [\href{https://arxiv.org/abs/1310.5150}{{\ttfamily 1310.5150}}].

\bibitem{Chen:2014gka}
Y.~Chen, R.~Harnik and R.~Vega-Morales, \emph{{Probing the Higgs Couplings to
  Photons in $h\to 4 \ell$ at the LHC}},
  \href{https://doi.org/10.1103/PhysRevLett.113.191801}{\emph{Phys. Rev. Lett.}
  {\bfseries 113} (2014) 191801},
  [\href{https://arxiv.org/abs/1404.1336}{{\ttfamily 1404.1336}}].

\bibitem{ATLAS:2015yrd}
{\scshape ATLAS} collaboration, G.~Aad et~al., \emph{{Constraints on
  non-Standard Model Higgs boson interactions in an effective Lagrangian using
  differential cross sections measured in the $H \rightarrow \gamma\gamma$
  decay channel at $\sqrt{s} = 8$TeV with the ATLAS detector}},
  \href{https://doi.org/10.1016/j.physletb.2015.11.071}{\emph{Phys. Lett. B}
  {\bfseries 753} (2016) 69--85},
  [\href{https://arxiv.org/abs/1508.02507}{{\ttfamily 1508.02507}}].

\bibitem{Voloshin:2012tv}
M.~B. Voloshin, \emph{{CP Violation in Higgs Diphoton Decay in Models with
  Vectorlike Heavy Fermions}},
  \href{https://doi.org/10.1103/PhysRevD.86.093016}{\emph{Phys. Rev. D}
  {\bfseries 86} (2012) 093016},
  [\href{https://arxiv.org/abs/1208.4303}{{\ttfamily 1208.4303}}].

\bibitem{Altmannshofer:2013zba}
W.~Altmannshofer, M.~Bauer and M.~Carena, \emph{{Exotic Leptons: Higgs, Flavor
  and Collider Phenomenology}},
  \href{https://doi.org/10.1007/JHEP01(2014)060}{\emph{JHEP} {\bfseries 01}
  (2014) 060}, [\href{https://arxiv.org/abs/1308.1987}{{\ttfamily 1308.1987}}].

\bibitem{Chao:2014dpa}
W.~Chao and M.~J. Ramsey-Musolf, \emph{{Electroweak Baryogenesis, Electric
  Dipole Moments, and Higgs Diphoton Decays}},
  \href{https://doi.org/10.1007/JHEP10(2014)180}{\emph{JHEP} {\bfseries 10}
  (2014) 180}, [\href{https://arxiv.org/abs/1406.0517}{{\ttfamily 1406.0517}}].

\bibitem{Bizot:2015zaa}
N.~Bizot and M.~Frigerio, \emph{{Fermionic extensions of the Standard Model in
  light of the Higgs couplings}},
  \href{https://doi.org/10.1007/JHEP01(2016)036}{\emph{JHEP} {\bfseries 01}
  (2016) 036}, [\href{https://arxiv.org/abs/1508.01645}{{\ttfamily
  1508.01645}}].

\bibitem{Egana-Ugrinovic:2018roi}
D.~Egana-Ugrinovic, M.~Low and J.~T. Ruderman, \emph{{Charged Fermions Below
  100 GeV}}, \href{https://doi.org/10.1007/JHEP05(2018)012}{\emph{JHEP}
  {\bfseries 05} (2018) 012},
  [\href{https://arxiv.org/abs/1801.05432}{{\ttfamily 1801.05432}}].

\bibitem{FileviezPerez:2011pt}
P.~Fileviez~Perez and M.~B. Wise, \emph{{Breaking Local Baryon and Lepton
  Number at the TeV Scale}},
  \href{https://doi.org/10.1007/JHEP08(2011)068}{\emph{JHEP} {\bfseries 08}
  (2011) 068}, [\href{https://arxiv.org/abs/1106.0343}{{\ttfamily 1106.0343}}].

\bibitem{Patel:2015tea}
H.~H. Patel, \emph{{Package-X: A Mathematica package for the analytic
  calculation of one-loop integrals}},
  \href{https://doi.org/10.1016/j.cpc.2015.08.017}{\emph{Comput. Phys. Commun.}
  {\bfseries 197} (2015) 276--290},
  [\href{https://arxiv.org/abs/1503.01469}{{\ttfamily 1503.01469}}].

\bibitem{Perez:2020baq}
P.~Fileviez~Perez, C.~Murgui and A.~D. Plascencia, \emph{{Baryonic Higgs and
  Dark Matter}}, \href{https://doi.org/10.1007/JHEP02(2021)163}{\emph{JHEP}
  {\bfseries 02} (2021) 163},
  [\href{https://arxiv.org/abs/2012.06599}{{\ttfamily 2012.06599}}].

\bibitem{ACMEIII}
J.~Doyle, \emph{Search for the electric dipole moment of the electron with
  thorium monoxide - the acme experiment},  2016,
  \href{http://online.kitp.ucsb.edu/online/nuclear\_c16/doyle/}{http://online.kitp.ucsb.edu/online/nuclear\_c16/doyle/}.

\bibitem{Arkani-Hamed:2004zhs}
N.~Arkani-Hamed, S.~Dimopoulos, G.~F. Giudice and A.~Romanino, \emph{{Aspects
  of split supersymmetry}},
  \href{https://doi.org/10.1016/j.nuclphysb.2004.12.026}{\emph{Nucl. Phys. B}
  {\bfseries 709} (2005) 3--46},
  [\href{https://arxiv.org/abs/hep-ph/0409232}{{\ttfamily hep-ph/0409232}}].

\bibitem{Carena:2018cjh}
M.~Carena, M.~Quirós and Y.~Zhang, \emph{{Electroweak Baryogenesis from
  Dark-Sector CP Violation}},
  \href{https://doi.org/10.1103/PhysRevLett.122.201802}{\emph{Phys. Rev. Lett.}
  {\bfseries 122} (2019) 201802},
  [\href{https://arxiv.org/abs/1811.09719}{{\ttfamily 1811.09719}}].

\bibitem{Carena:2019xrr}
M.~Carena, M.~Quirós and Y.~Zhang, \emph{{Dark CP violation and gauged lepton
  or baryon number for electroweak baryogenesis}},
  \href{https://doi.org/10.1103/PhysRevD.101.055014}{\emph{Phys. Rev. D}
  {\bfseries 101} (2020) 055014},
  [\href{https://arxiv.org/abs/1908.04818}{{\ttfamily 1908.04818}}].

\bibitem{Perez:2021udy}
P.~F. Perez, C.~Murgui and A.~D. Plascencia, \emph{{Baryogenesis via
  leptogenesis: Spontaneous B and L violation}},
  \href{https://doi.org/10.1103/PhysRevD.104.055007}{\emph{Phys. Rev. D}
  {\bfseries 104} (2021) 055007},
  [\href{https://arxiv.org/abs/2103.13397}{{\ttfamily 2103.13397}}].

\bibitem{Bennett:2008dy}
{\scshape Muon (g-2)} collaboration, G.~W. Bennett et~al., \emph{{An Improved
  Limit on the Muon Electric Dipole Moment}},
  \href{https://doi.org/10.1103/PhysRevD.80.052008}{\emph{Phys. Rev. D}
  {\bfseries 80} (2009) 052008},
  [\href{https://arxiv.org/abs/0811.1207}{{\ttfamily 0811.1207}}].

\bibitem{Nakai:2016atk}
Y.~Nakai and M.~Reece, \emph{{Electric Dipole Moments in Natural
  Supersymmetry}}, \href{https://doi.org/10.1007/JHEP08(2017)031}{\emph{JHEP}
  {\bfseries 08} (2017) 031},
  [\href{https://arxiv.org/abs/1612.08090}{{\ttfamily 1612.08090}}].

\end{thebibliography}\endgroup
\end{document}